\documentclass[twocolumn,aps,amssymb,footinbib,floatfix,pre,10pt,longbibliography]{revtex4-2}
\usepackage{notoccite}
\usepackage{amssymb}
\usepackage{graphicx}  
\usepackage{amsmath}
\usepackage{nicefrac, xfrac}
\usepackage{mathrsfs}
\usepackage{times}
\usepackage{color}
\usepackage{subfigure}
\usepackage{bbold}
\usepackage{bm}
\usepackage{mathtools}
\usepackage{notes2bib}

\usepackage[usenames,dvipsnames]{xcolor}
\usepackage[%
   colorlinks=true,
   pdfborder={0 0 0},
   linkcolor=violet,
   citecolor=MidnightBlue
]{hyperref}
\hypersetup{colorlinks=true, citecolor=MidnightBlue, urlcolor=MidnightBlue, linkcolor=teal}

\usepackage{yfonts}

\newcommand{\rr}{\mathbf{r}}
\newcommand{\ff}{\bm{f}}
\newcommand{\RR}{\mathbf{R}_p}
\newcommand{\xx}{\hat{\bm{x}}}
\newcommand{\vv}{\bm{v}}
\newcommand{\uu}{\mathcal{U}}
\newcommand{\nn}{\hat{\mathbf{n}}}
\newcommand{\pwr}{\mathcal{P}}
\newcommand{\dd}{{\text{d}}}
\newcommand{\DD}{{\text{D}}_t}
\newcommand{\vp}{{\bf{V}}_p}

\newcommand{\vl}{V_{\ell}}
\newcommand{\vaf}{V_{\text{AF}}}
\newcommand{\bep}{\bm{\varepsilon}}
\newcommand{\bsg}[1]{{\bm{\Sigma}}^{\text{#1}}}
\newcommand{\tauk}{\tau_K}
\newcommand{\xdw}{X_{\text{DW}}}
\newcommand{\vdw}{V_{\text{DW}}}
\newcommand{\D}[1]{\delta{#1}}

\newcommand{\mrm}[1]{\mathit{#1}}

\newcommand{\PRLsep}{\noindent\makebox[\linewidth]{\resizebox{0.6\linewidth}{1pt}{$\bullet$}}\bigskip}

\newcommand{\db}{\text{d}\hspace*{-0.15em}\bar{}\hspace*{0.15em}}

\usepackage[T1]{fontenc}  		    
\usepackage{textcomp}     		    

\begin{document}
        
\begin{center}
    \begin{abstract}
        \noindent Active materials form a class of far-from-equilibrium systems that are driven internally and exhibit self-organization, which can be harnessed to perform mechanical work. Inspired by experiments on synthetic active polymer networks, in this paper we examine limits of work extraction from an active viscoelastic medium by analyzing the transport of a particle. The active viscoelastic material possesses an equilibrium density where the active and passive forces are balanced out. In a one-dimensional system, a gliding activation front (AF) that converts a passive to an active medium, provides active energy at a constant rate, which is injected into the system at one end and propagates to the other. {We demonstrate that there exists a maximum velocity of the AF, above which the activated region fails to deliver the transport power. We hypothesize, and intuitively argue based on the limit cases, that the feasibility and the velocity of transport of the particle can be interpreted in terms of the velocity of an equilibration Domain Wall of the field, which is set by two  parameters: (i) a measure of activity, and (ii) the viscoelastic timescale. The phase diagram is divided into ``Transport'' and ``No-Transport'' sectors, namely for any pair of the two parameters, there exists a threshold velocity of the AF above which the particle transport becomes impossible. Constructing the phase diagram we find that: (1) there are regions of the phase diagram for which the threshold velocity of the AF diverges, (2) larger viscoelastic timescale makes the transport region more accessible, and also increases the transport velocity therein. Furthermore, we find that increasing the velocity of AF, results in larger extracted power but smaller transport coefficient; defined as the ratio of the transport velocity and that of the AF.} Our model provides a framework for understanding the energetics of transport phenomena in biology, and designing efficient mechanisms of transport in synthetic active materials.
    \end{abstract}
\end{center}

\title{Theoretical Limits of Energy Extraction in Active Fluids}

\author{Shahriar Shadkhoo}
\email{\textcolor{MidnightBlue}{shahriar@caltech.edu}}
\author{Matt Thomson}
\affiliation{California Institute of Technology, Pasadena, CA, 91125}

\maketitle

\section{Introduction}
For more than a century the problem of energy extraction from non-equilibrium systems has been a subject of extensive research, with significant implications in fundamental physics and engineering. Thought experiments like Maxwell's demon and the generalizations thereof have established deep connections between information gain and work extraction \cite{maruyama2009colloquium}. Such studies have also advanced our understanding of the optimal protocols of energy extraction \cite{liphardt2002equilibrium,wang2002experimental,carberry2004fluctuations,thorn2008experimental,lopez2008realization,toyabe2010experimental,sagawa2010generalized,gupta2023efficient,kumari2023stochastic}. The role of feedback loops in the process of information-to-energy conversion was highlighted in an early instantiation of the Maxwell's demon suggested by Szil\'{a}rd \cite{szilard1964decrease}. {This paradigm of energy extraction at the microscopic scales leverages information to rectify the fluctuations in the desired direction, hence decreasing entropy.}

Energy can also be extracted at macroscopic scales from systems that are externally supplied with energy---usually at the boundaries. The injected energy excites the dynamical modes of the system which may subsequently perform work. Active materials constitute a class of far-from-equilibrium systems that are driven internally via injection of energy at microscopic scales, and exhibit self-organization---a common-place phenomenon in biology. In spite of its prevalence, the mechanisms of energy propagation and conversion in active systems is yet to be elucidated. Unraveling such mechanisms would provide insights into the efficiency of biological pathways and the barriers that impede the conversion of active energy into mechanical work \cite{foster2023dissipation}. Besides natural realizations, synthetic active systems allow us to program novel states of matter through harnessing the power of self-organization, and designing optimal control protocols \cite{theillard2017geometric,ross2019controlling,chandrakar2020confinement,norton2020optimal,ziepke2022multi,wang2022forces,najma2022competing,shankar2022optimal,hecht2022active,lamtyugina2022thermodynamic,derivaux2023active}. 

\begin{figure}[b]
    \centering
        \centerline{\includegraphics[width=8.5cm]{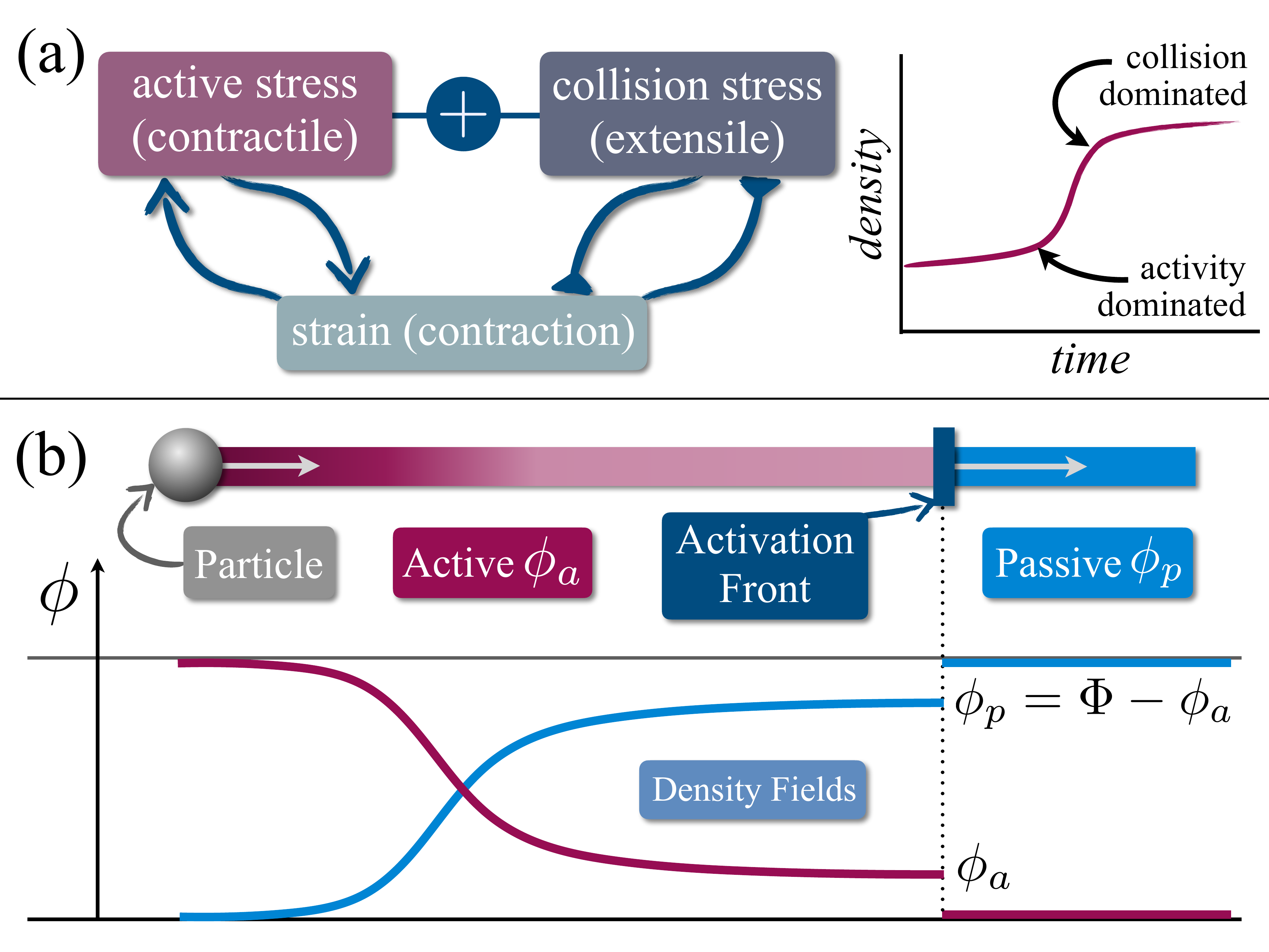}}
            \caption{(a) The positive and negative feedback loops between strain and the two competing contractile and extensile components of stress, which originate from the active and passive collisions, respectively. (b) Transport of a particle coupled to an active string. The activation front generates stress gradient and pulls on the particle. The densities of the passive and active components are denoted by $\phi_p$ and $\phi_a$, respectively, the sum of which equals $\Phi$, a constant.}
        \label{mt.fig.schematic}
    \vspace{-3mm}
\end{figure}

In this article we propose a model of an active system with a built-in mechanical feedback which can be utilized to extract energy. The feedback mechanism operates on the basis of activity-induced contraction in viscoelastic active materials, which in turn amplifies local activity and the resultant contraction; Fig. (\ref{mt.fig.schematic}a). Coupling a particle to the active medium, and studying the transport of the particle, we investigate the limits of energy extraction from the \emph{boundary} of scalar viscoelastic active systems; see Fig. (\ref{mt.fig.schematic}b). 

The limits of energy extraction are determined by the properties of the system at hand. Here we consider a continuum model, unlike most studies where the stochastic thermodynamics of the microscopic degrees of freedom is considered \cite{pietzonka2017entropy,pietzonka2019autonomous,malgaretti2022szilard,ro2022model,datta2022second,cocconi2023optimal}. A continuum description of viscoelasticity comprise an elastic and a viscous component expressed in terms of the strain and stress fields. The elasticity and viscosity together define a timescale, that characterizes the timescale over which the dominant mode transitions from viscous to elastic, or vice versa. In our model---inspired by the dynamics of active bio-polymer networks---activity, which introduces the source of the to-be-extracted energy, is manifested in the form of density-dependent contractile stresses (see Sec. (\ref{mt.sec.model})). Stress gradients provide forces that can potentially perform work on a second system that is coupled to the active one. In a system with statically defined boundaries, i.e. no material exchange, the energy extraction (and motion) cannot be sustained in the steady-state sense. The active energy excites the dynamics, which is subsequently damped out towards an equilibrium state, in which the active energy is constantly converted into the elastic energy of the equilibrium state. Should the system be provided with constant input of fresh active energy, however, the active energy can potentially provide the power required for sustained dynamics.

Steady-state transport of an external particle in a viscous medium exemplifies the extraction of energy. Transport phenomena require net (thermodynamic) forces like pressure gradients. In isolated passive systems, transport of the constituents is driven by minimization of the free energy towards the thermodynamic equilibrium \cite{chaikin1995principles}. Active materials utilized bulk energy to perform work at large scales \cite{de2015introduction}. Important examples of active transport in biology include cytoplasmic transport of proteins and ions \cite{suh2003efficient,ross2008cargo,goswami2008nanoclusters,yeomans2017nature,guo2021play,saha2021maximizing,foster2023dissipation}. The study of active transport has been mostly---with a few exceptions e.g. in \cite{schildknecht2021phenomenological,pietzonka2017entropy}---limited to self-propelled particles in a passive medium \cite{chopra2022geometric,cagnetta2020efficiency,reichhardt2014active, marchetti2013hydrodynamics,neri2013modeling,costanzo2012transport}. An alternative scenario is for the particle to be coupled to a reservoir of active constituents that provides the transport energy. 

Here we aim to understand the properties of the active system that determine the limits of energy extraction. Our model is motivated by an experimental work in which motility is induced in a mixture of motor proteins and microtubule filaments. The motor proteins are activated upon shining light, form cross-links and walk along the filaments \cite{ross2019controlling,liu2024force}. A microtubule dense aster resembling the external rigid particle, is transported as the activation front sweeps across the system at a fixed velocity. An upper bound on the transport velocity is observed, when the velocity of the activation front exceeds a threshold value. Even though in the previous experiments the mechanical properties of the active networks (such as elasticity, viscosity and the magnitude of active stress) vary over narrow ranges of parameters, the synthetic nature of such systems provides the possibility of exploring their behavior beyond the currently available ones. Therefore, extrapolating the dynamical behavior to broader ranges of parameters allow us to design systems with higher efficiencies.

In our system, energy is provided by a gliding activation front (AF) that activates an otherwise passive medium; like a combine harvester machine. The progression of the AF leaves behind a trail of activated material which evolves under its active stress, and induces a stress gradient that drives the particle; see Fig. (\ref{mt.fig.schematic}b). Taking the experimental work as our phenomenology, we formulate a model that allows us to explore the phase diagram of the system. Our model involves several parameters (see Table \ref{sm.table1}), among which we choose the following ones to study: (1) a measure of activity, (2) the viscoelastic timescale, (3) the velocity of the activation front, and (4) the coupling to the particle. Our theory provides insight into the mechanism of transport within the active system and also reveals how an upper bound on transport velocity emerges within a contractile active system. {Note that in our setting, the particle is not a probe that extracts energy from the active medium, but instead interacts with both active and passive components, and ``measures'' the amount of extracted energy expended by the active medium to transport the particle against the drag force. This is distinct from the viscous dissipation of the material that occurs even in the absence of the particle. We show that the viscoelastic timescale which characterizes the propagation velocity of the density wave, is a key determinant of the feasibility and the efficiency of transport. In the context of active polymer networks, this timescale is determined by the rate of cross-linking filaments and the consequent increase of the rigidity over time.}

A key difference between our setting and conventional active systems is that here, the active and passive stresses can cancel out one another, where the system ceases to flow. Therefore, even though active energy is required to be constantly injected \emph{locally} for maintaining the stress (force dipoles) in the contracted state, a persistent flow requires nonzero divergence of the stress, which occurs when fresh active {material} is introduced to the system.

\begin{table}{\label{sm.table1}}
    \begin{tabular}{ |c||c| }
         \hline
         \textbf{Parameter} & \textbf{Definition} \\
         \hline\hline
        $\alpha_{1,2}$ & Coefficients of stress vs. density\\
         \hline
        $\eta$ & Viscosity of the fluid component\\
         \hline
        $\kappa$ & Elasticity of the rigid component\\
         \hline
        {$\tau_K$}{ \scriptsize$={\kappa}^{-1}$} & Kelvin-Voigt viscoelastic time\\
         \hline
        $\gamma$ & Drag coefficient of the field\\
         \hline         
        $\phi_i$ & Initial density of activated region \\
         \hline
        $\vaf$ & Velocity of the Activation Front \\
         \hline
        $g_0$ & Field-particle coupling constant\\
         \hline
        $\Gamma$ & Drag coefficient of the particle\\
         \hline\hline
         {\textbf{Variable}} & {\textbf{Definition}} \\
         \hline\hline
        $\bsg{}\,,\,\sigma$ & Stress field of active region\\
         \hline
        $\bm{\mathcal E}\,,\,\bep$ & Strain field of active region\\
         \hline
        $\phi_a$ & Active density field\\
         \hline
        $\phi_e$ & Equilibrium density\\
         \hline
        $\phi_b$ & Boundary density\\
         \hline         
        $v_b$ & Boundary velocity \\
         \hline
        $\vdw$ & Velocity of the Domain Wall \\
         \hline
    \end{tabular}
    \caption{Model parameters and variables with their definitions.}
\end{table}

\section{Model}{\label{mt.sec.model}}
The active system we adopt to study comprises a mixture of active and passive components, the total density of which is initially, and remains, uniform in space: $\Phi = \phi_a + \phi_p$, where $\phi_{a,p}$ denote the corresponding densities of active and passive components. To put our model in context we consider a mixture where the passive component is a solution of free-floating filaments and cross-linkers, that would assemble into an active network (gelation) upon ``activation'' which entails recruiting active cross-linkers \cite{prost2015active}. Active networks of filaments can be realized naturally in biological systems, or artificially (e.g. using iLID technology \cite{guntas2015engineering,ross2019controlling,liu2024force}), and are driven out of equilibrium by active cross-linkers that generate stress by walking along the filaments. The magnitude of the active stress is controlled by a parameter that depends on the density of active cross-linkers \cite{foster2015active}. We find it necessary to mention that while we use active networks as a synthetic system, to which our model is applicable, some aspects of the model are independent of the details of such systems. Therefore we define phenomenological parameters that would have various microscopic interpretations depending on the system.

\subsection{Active Viscoelasticity}{\label{mt.sec.activeviscoelasticity}}
In our model, activation is defined as a process that endows the passive component with the capacity of forming an active gel of initial density $\phi_i$. The density in the case of polar filaments represents the density of filaments' plus (or minus) ends in the cross-linked network (see Ref. \cite{foster2015active}). While activation is instantaneous, gelation takes place over a characteristic timescale of the system, suggesting viscoelastic behavior \cite{gittes1997microscopic,deng2006fast,mackintosh2008nonequilibrium,prost2015active}. The specific model of viscoelasticity to be adopted depends on the timescales in the phenomenology of interest. There exist two minimal models of viscoelasticity, that can be represented by means of a spring and a dashpot, that are arranged either in series or parallel configurations. The series configuration, called Maxwell's model describes high-frequency elastic followed by low-frequency viscous behavior, where as the parallel configuration, called Kelvin-Voigt model represents the opposite behavior. The two models (Maxwell and Kelvin-Voigt) both describe dynamics that transition between elasticity- and viscosity-dominated modes. While the response of the preformed polymer networks is primarily described in terms of Maxwell's viscoelasticity \cite{prost2015active}, the response of actively assembling (as opposed to preformed) networks transitions from viscous to elastic in time, and can be captured by Kelvin-Voigt model \cite{hurst2021intracellular}. The elastic to viscous transition of Maxwell's model in preformed networks, originates from breaking the cross-links under stress, which leads to relaxing internal stresses. On the other hand, a network that is undergoing cross-linking behaves like a fluid before the cross-linking reaches the limit of stress percolation where the network exhibits elastic behavior. As such, given that the activation front in our system ignites the cross-linking in a previously free-floating mixture of filaments, the short term behavior is dominated by viscosity and gradually transitions to elasticity; hence Kelvin-Voigt model.

In addition to the density field $\phi_a$, the active component is characterized by a displacement field $\bm u_a(\rr,t)$. Using Euler's notation for derivatives: $\DD f\equiv\frac{\text{D}f}{\text{D}t}$ the material time derivative, and $\partial_qf \equiv \frac{\partial f}{\partial q}$ the partial derivative with respect to variable $q$, the velocity and strain fields equal $\vv_a=\partial_t \bm u_a$, and $\bm{\mathcal{E}} = \nabla \bm u_a$. Together with compressible continuity equation $\partial_t \phi_a + \nabla \cdot (\phi_a \vv_a) = 0$, momentum conservation and the constitutive equation read:
\begin{subequations}{\label{mt.eq.constitutive}}
   \begin{gather}
       \DD(\phi_a\vv_a) = \nabla\cdot\bsg{} + \ff,{\label{mt.eq.newton2nd}}\\
       \begin{split}
           \bsg{} &= \eta\left(\tauk^{-1}+ \DD\right)\bm{\mathcal{E}} \\
           &= \kappa \,\bm{\mathcal{E}}+ \eta\, \DD\,\bm{\mathcal{E}}.
       \end{split}
   \end{gather}
\end{subequations}
Here $\bsg{}$ and $\bm{\mathcal{E}}$ represent stress and strain fields; $\eta$ and $\tauk$ are the viscosity and the Kelvin timescale, and $\kappa = \eta/\tauk$ denotes the rigidity of the elastic component. The body force density $\ff$ reads $\ff=-\nabla\cdot\bsg{int}-\gamma_0\phi_a\phi_p(v_a-v_p)$. The second term is the drag force between passive and active components that respects $\{a\}\leftrightarrow \{p\}$ exchange symmetries. Using the continuity equation $\phi_av_a + \phi_pv_p = 0$, we can show that the drag reduces to $-\gamma_0\Phi\phi_av_a\equiv-\gamma\phi_av_a$ (see Appendix Sec. (\ref{sm.sec.model})). The internal stress is the sum of contractile (active) and an extensile (passive) components. The latter originates from the collisions of the filaments, which guarantees stability against the collapse of the network under contractile stress. Using a virial-like expansion in terms of the density field we get:
\begin{align}{\label{mt.eq.stress}}
    \bsg{int} = \sigma^{\text{int}}(\phi_a) \, \mathbb{I} &= -\alpha_1 \phi^{\phantom{}}_a\mathbb{I} + \alpha_2 \phi_a^2\mathbb{I} + \mathcal{O}(\phi_a^3)\\
    &= -\alpha_1\phi_a(1-{\phi_a}/{\phi_0})+ \mathcal{O}(\phi_a^3),
\end{align}
where $\alpha_1,\alpha_2>0$ are constants, and $\mathbb{I}$ is the identity matrix. To second order in the density field, the stress can be written as: $\sigma^{\text{int}}=-\alpha_1\phi_a\,(1-\phi_a/\phi_0)$, where $\phi_0=\alpha_1/\alpha_2$, is the density at which the active and passive densities cancel out one another; but is in general distinct from equilibrium density for finite Kelvin timescales $\tauk$; see below. Since we are interested in one-dimensional transport, the tensors are reduced to scalar fields $\sigma$ and $\bep$ which represent $\bsg{}$ and $\bm{\mathcal{E}}$, respectively.

The signature of activity can be seen by noting that the contractile stress and density form a positive \emph{feedback loop} against the conventional equilibration in passive systems---except for gravitational systems. The positive feedback holds for $\phi_a\leq\phi_0/2$, above which the effect of passive stress dominates the contractile term. At equilibrium, the internal stress is given by $\sigma^{\text{int}}(\phi_e) = \kappa \bep_e$. In Appendix Sec. (\ref{sm.sec.model}), we show that a dimensionless number, $\nu = \alpha_1^2\tauk/\eta\alpha_2 = \alpha_1^2/\kappa\alpha_2$, controls the equilibrium density which varies between $\phi_i<\phi_e\leq\phi_0$ for $\alpha_1\geq \alpha_2\phi_i$ (the condition required for contraction in the first place). In the limit of minuscule elasticity $\kappa\to0$, the asymptotic solution reads: $\phi_e = \phi_0 - (\phi_0-\phi_i)\,\nu^{-1} + \mathcal O(\nu^{-2})$. In the opposite limit $\nu\to 0$, we get: $\phi_e=\phi_i+(\phi_i^2/\phi_0)\left(1-{\phi_i}/{\phi_0}\right)\nu+\mathcal O(\nu^{2})$; also see Appendix Fig. (\ref{sm.fig.straineq}) for numerical solutions.

Finally, the dissipation of energy occurs through the viscosity of the active component, and the drag against the background medium. While the former is a characteristic of the active material, the latter is medium-specific. Thus we deliberately neglect the drag force to focus on the material properties. In Sec. (\ref{mt.sec.discuss}) below, and in Appendix (\ref{sm.sec.drag}), we discuss the effect of nonzero drag.

\subsection{Field-Particle Interaction}
The total force experienced by a particle embedded in the solutions depends on passive and active pressures. To linear order, the total pressure reads $P=P_a+P_p$, where $P_{a,p}=g_{a,p}\phi_{a,p}$, and $g_{a,p}$ are the coupling constants. The driving force equals: $F = \ell\, \nabla P$, where $\ell$ is the particle's linear size. Given that $\phi_a+\phi_p=\Phi$ is constant, we write the pressure gradient as $\nabla P=(g_a-g_p)\nabla\phi_a=g_0\nabla\phi_a$, where $g_0=g_a-g_p$. Note that $g_0$ and $\alpha_1$ are of the same dimensions. In the experiments of light-induced activity in the cross-linked networks, high-density localized asters that are isolated from the passive medium, can play the role of a rigid particle. With this analogy in mind, we choose $g_0>0$; meaning that the particle is pulled by the newly activated regions. The case of $g_0<0$ is briefly discussed below and in Appendix Sec. (\ref{sm.sec.ParticleInteraction}).

\begin{figure}
    \centering
        \centerline{\includegraphics[width=9 cm]{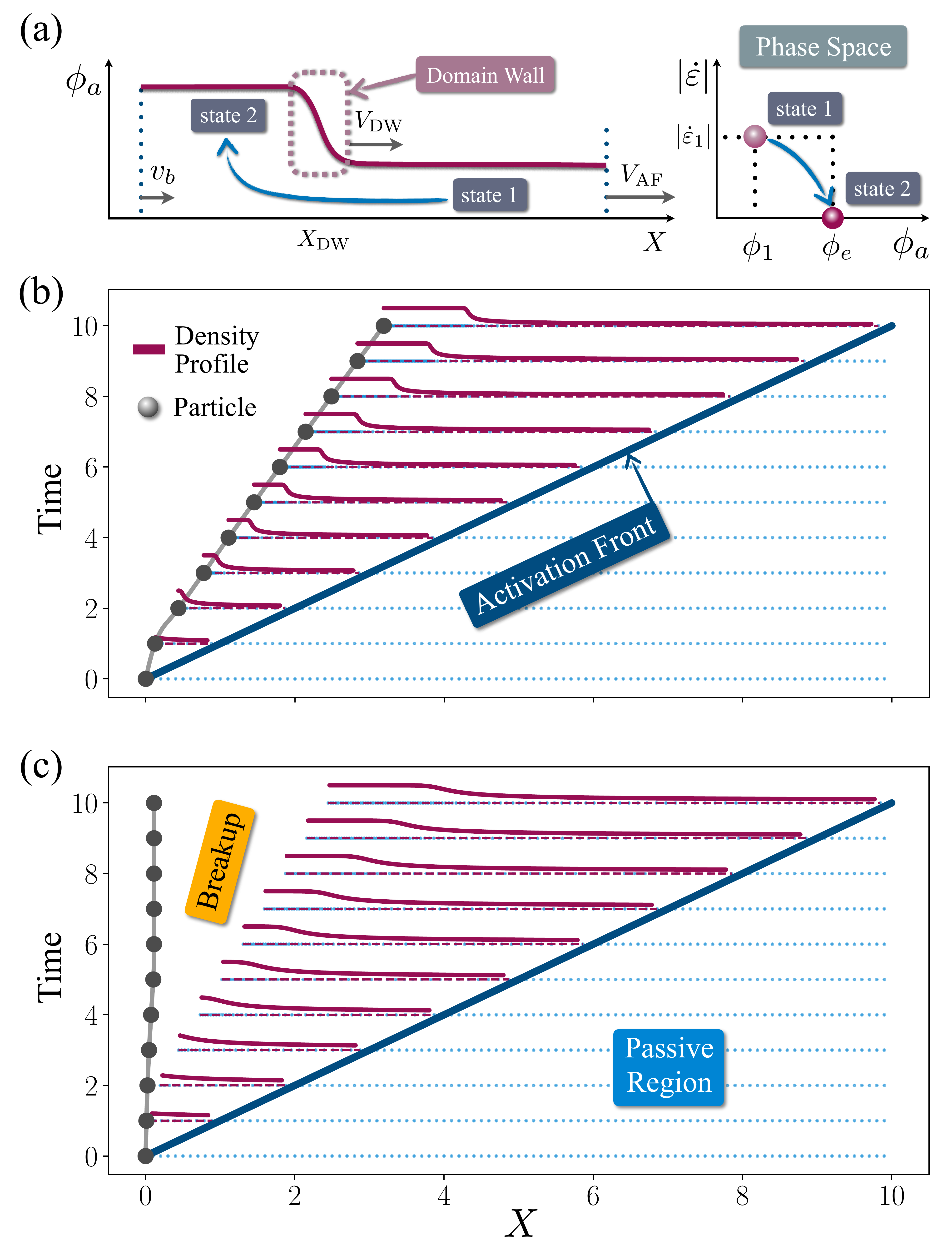}}
            \caption{(a) A schematic of the transport of a particle, as the activation front (dark blue line in (b) and (c)) progresses in time and activates the passive region. The newly activated matter transitions state 1 characterized by ($\phi_1, \dot{\bep}_1$), to the equilibrium state 2 with ($\phi_2=\phi_e, \dot{\bep}_2=0$). The two states are separated by the DW moving with velocity $\vdw$. The left boundary moves with $v_b$. The density profiles and the position of the particle are shown for two sets of parameters: (b) large enough coupling for the activation front to drag the particle, hence steady transport; and (c) small coupling the particle is detached and the transport fails.}{\label{mt.fig.motion}}
    \vspace{-3mm}
\end{figure}

The motion of the particle is resisted by a drag force, which can be shown to be equal to $-\Gamma \vp$, with $\Gamma$ the effective drag coefficient (see Appendix Sec. (\ref{sm.sec.ParticleInteraction})). In the overdamped limit the equation of motion reads: $\Gamma\vp = g_0\ell\,\nabla\phi_a$. In order to focus on the properties of the active system regardless of its specific coupling to the particle, here we assume that the coupling to the particle is negligible compared to the field's internal interactions: $g_0\ll\alpha_1$; namely we neglect the effect of the particle on the field. The potential field to which the particle is exposed is obtained by $\uu=-g_0\ell\int^x\dd x\nabla\phi_a = -g_0\ell\phi_a$, which is not influenced by the particle---similar to the system studied in \cite{venturelli2023stochastic}. 

An interesting point is that the feasibility of transport depends on the initial condition. If the \emph{initial} positions of the AF and the particle coincide, the transport is only possible for $g_0>0$. If the particle is introduced to an already activated region (with density gradient), the transport can \emph{also} occur for $g_0<0$. In the latter case, the particle is trapped at the position of the moving density domain wall which carries a negative density gradient (see Fig. (\ref{mt.fig.motion}a) for the definition of the domain wall). The two scenarios introduce two potential wells that could trap and drag the particle along, though with opposite curvatures of the density profile. Here we are focusing on the case of $g_0>0$, which results in the particle being trapped at the boundary of the system. 

Throughout the paper we set $\alpha_2$ (as opposed to $\phi_0$) equal to unity; thus $\phi_0 = \alpha_1$. The reason behind this choice is that in real systems, the passive (excluded volume) pressure, characterized by $\alpha_2$, is independent of the activity. Furthermore, even though the field's viscosity $\eta$, the particle's drag coefficient $\Gamma$, and the initial density $\phi_i$, are kept around in our calculations, the numerical results are obtained by setting them equal to unity.

\section{Active Transport}
The hypothesis we follow is that by continuously activating the material in a specific direction with velocity $\vaf$, contractile stress provides the power required to move the particle. Given the uniformity of the passive region, $\vaf$ is proportional to, and can be interpreted as the \emph{rate of active energy} injected into the system. We numerically explore the phase diagrams of transport spanned by $(\alpha_1,\vaf,g_0)$; see Fig. (\ref{mt.fig.PD}). We elaborate on the numerical methods in Appendix Sec. (\ref{sm.sec.numerics}). In specific limits we also derive analytical solutions; Sec. (\ref{mt.sec.largeVaf}).

To gain intuition into the dynamics we use a discrete picture. Moving the activation front in $+\xx$ direction, the activated segments start contracting and their densities increase. The time difference between the activation of successive segments creates a negative density gradient. Denoting the positions of the particle and that of the boundary of the active region by $X_p(t)$ and $x_b(t)$, respectively, we argue that should the particle be initially at the same position as the activation front (i.e. $X_p(t=0)=x_b(t=0)$), the particle can \textit{potentially} stay trapped at the boundary of the active region. The reason is as follows: the potential in the immediate vicinity of the boundary of the active region resembles a step function $\sim \Theta(x-x_b(t))$. For $x>x_b$, the negative density gradient yields negative force, pushing the particle back towards the boundary, rendering it a stable minimum. In the long time limit the density gradient at $x\to x_b^{+}$ approaches $\nabla\phi\to 0^{-}$, and $\phi_b$ approaches the equilibrium value $\phi_e$. {As shown previously, the dimensionless parameter $\nu \propto \alpha_1^2/\kappa$ controls the boundary's equilibrium density and thus the force exerted on the particle.} The force exerted on the particle, and its velocity, are given by $g_0\phi_e$, and $V_p = g_0\phi_e/\Gamma$, respectively. This is referred to as the steady state of the particle---distinct from the equilibrium of an isolated active field. Below we derive the conditions on the field's parameters required for the particle to stay trapped in the boundary potential well.

\begin{figure}
    \centering
        \centerline{\includegraphics[width=8cm]{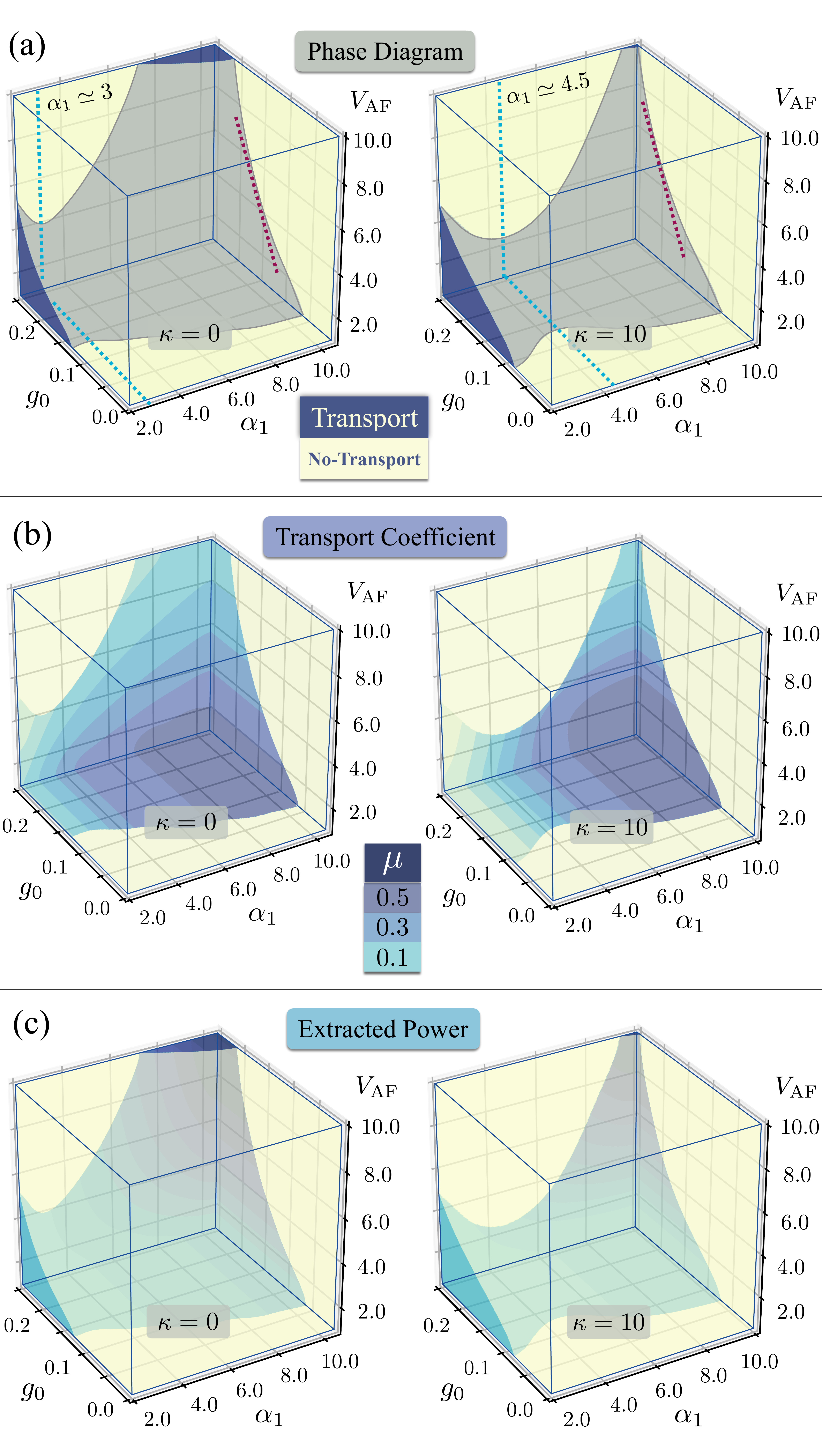}}
            \caption{Three dimensional phase diagram of the transport in the $(\alpha,g_0,\vaf)$ space for infinite viscoelastic timescale $\kappa=0$ (left), and finite timescale, $\kappa=10$ (right). (a) The navy blue and yellow sectors represent the Transport and No-Transport regions, respectively. For fixed values of $\alpha_1$ the boundary in $g_0-\vaf$ planes make convex curves (red dotted lines). (b) Contours of the transport coefficient shown in the transport phase. showing that the maximum ratio $\mu$ is achieved for larger and smaller values of $\alpha_1$ and $\vaf$, respectively. {(c) The transport power defined as $\pwr_t = \Gamma V_p^2$, which increases (nonlinearly) with $\alpha_1$ and $\vaf$.}} 
        \label{mt.fig.PD}
    \vspace{-3mm}
\end{figure}

\subsection{Phase Diagram and Transport Transition}
The active energy and stress, generated at the AF, propagate across the system via a traveling domain wall (DW) of density modulation (see the schematic in Fig. (\ref{mt.fig.motion}a)). We speculate that there exists a maximum velocity of the activation front $\vaf$, beyond which the particle escapes the step potential. For a trapped particle we have $X_p(t) = x_b(t)$, and $V_p = v_b$. {Using force balance we get $g_0\phi_b\geq \Gamma V_p$ as the transport condition. If the force exerted by the field is smaller than the drag, the particle escapes the potential well; on the other hand if it is larger than the drag, the particle is constantly pushed forward momentarily, as a result of which the particle experiences a positive potential gradient and returns to the boundary. More precisely by regularizing the step potential with a smooth function of infinitesimally small width, the potential gradient varies continuously---over an interval---between a large negative value and a positive value equal to $\nabla\phi_b$. The transport condition above guarantees that the potential gradient equals the drag force at some point along this interval.} Equivalently, we have $\displaystyle\lim_{t\to\infty}(v_b/\phi_b) \leq g_0/\Gamma$. Since $v_b<\vaf$ and $\phi_b=\phi_e$, a sufficient (but not necessary) trivial condition on $\vaf$ would be $\vaf \leq g_0\phi_e/\Gamma$. The actual upper bound is obtained by noting that in the {long time} limit, the boundary density and velocity read $\displaystyle\lim_{t\to\infty}\phi_b=\phi_e$, and $\displaystyle\lim_{t\to\infty}v_b=\mu\vaf$, where $\mu(\alpha,\vaf)<1$, is a concave function of $\alpha_1$ and a convex function of $\vaf$; see Appendix Fig. (\ref{sm.fig.mu}). As a matter of fact, we show that the boundary velocity $v_b$ approaches a constant value for infinitely large $\vaf$; hence the convexity.

For finite values of $\vaf$, we cannot analytically solve for $\mu(\alpha,\vaf)$ and $v_b$. However, calculating the extracted power does not require that information. In the limit of purely viscous medium ($\tauk\to\infty$) the equilibrium density of the boundary equals $\phi_b=\phi_0=\alpha_1/\alpha_2$ (see Appendix Fig. (\ref{sm.fig.straineq})), thus $\mu^*\vaf^*$ satisfies $\lim_{\tauk\to\infty}(\mu^*\vaf^*) = \frac{\alpha_1g_0}{\alpha_2\Gamma}$. The dependence of $\vaf^*$ on $\tauk$ are illustrated in Fig. (\ref{mt.fig.PD}a) and elaborated on in Appendix Sec. (\ref{sm.sec.transport}) and in Appendix Fig. (\ref{sm.fig.kk_series}). Increasing $\kappa$ (decreasing $\tauk$) results in smaller boundary density, which pushes the phase boundary towards smaller $\vaf$ and larger $(\alpha_1,g_0)$ (except at small $\alpha_1$ which is discussed separately below). In the limit $\tauk\to0$, we have $\phi_b \to \phi_i$. The threshold velocity satisfies: $\displaystyle\lim_{\tauk\to0}(\mu^*\vaf^*)=\phi_ig_0/\Gamma$.

The transport power equals $\pwr_t=\Gamma V_p^2$. The power delivered by the field at its boundary is $g_0\phi_bv_b$ which is maximized for $\max(\phi_bv_b)$, subject to the transport condition $v_b/\phi_b \leq g_0/\Gamma$. Therefore the power satisfies $\pwr_t\leq (g_0/\Gamma)\phi_b^2$, where $\phi_b=\phi_e$ depends on $\alpha_1,\tauk$ as discussed in Sec. (\ref{mt.sec.activeviscoelasticity}) and shown in Appendix Fig. (\ref{sm.fig.straineq}). Therefore in general we have:
\begin{equation}
    \pwr_t^{\max}\big|_{(\alpha_1,\tauk)} = {\Gamma}^{-1}g_0^2\phi_e^2(\alpha_1,\tauk).
\end{equation}
The maximum equilibrium density on the r.h.s. equals $\max(\phi_e) = \alpha_1/\alpha_2$ which is achieved for $\tauk\to\infty$ (zero elasticity). Therefore the maximum extracted power equals:
\begin{equation}{\label{eq.pt}}
    \max_{\tauk} \pwr^{\max}_t = \frac{\alpha_1^2g_0^2}{\Gamma\alpha_2^2}.
\end{equation}
The minimum of the maximum power is, on the other hand, achieved for $\tauk\to0$ and equals $\min_{\tauk}\pwr^{\max}_t = g_0^2\phi_i/\Gamma$. The transport power found in Eq. (\ref{eq.pt}), resembles the power required for a constant current in an Ohmic material $\pwr=V^2/R$; where the resistance $R$ plays the role of the particle's drag coefficient $\Gamma$, and the potential difference is given by $V\equiv \alpha_1g_0/\alpha_2$. More specifically, $g_0$ is equivalent to the carriers' charges coupled to $\alpha_1/\alpha_2$ as the electromotive force.

\subsubsection{Numerical Results}
Using the numerical methods the details of which can be found in Appendix Sec. (\ref{sm.sec.numerics}), we construct the phase diagram in terms of the activity $\alpha_1$ and the coupling $g_0$, and for two different values of $\kappa = 0$ and $\kappa=10$ (or equivalently for $\tauk = 0.1$ and $\tauk\to\infty$). The phase diagram portrayed in Fig. (\ref{mt.fig.PD}a), separating Transport and No-Transport regimes, contains important information regarding the maximum (i.e. threshold) velocity of the activation front, $\vaf$, for which the transport is feasible. For small values of $\vaf$ the transport is feasible in broader ranges of the $\alpha_1-g_0$ subspace. At any value of $\alpha_1$, the projections of the transition region onto the $g_0-\vaf$ planes show that the phase boundary is a convex curve, which extrapolates down to trivial point of $(g_0,\vaf)=(0,0)$, (not shown in the figure). The convexity of the phase boundary curves at fixed $\alpha_1$'s suggests that, in terms of $g_0$, the threshold velocity $\vaf^*$ for transport grows super-linearly. In Sec. (\ref{mt.sec.largeVaf}), we see that $\vaf^*$ indeed diverges, under certain conditions.

In the transport sector of the phase diagram, the particle velocity equals the boundary velocity. As such, we define the transport coefficient: $V_p/\vaf = v_b/\vaf = \mu$; which is essentially identical to the function defined above, and vanishes in the No-Transport region. Figure (\ref{mt.fig.PD}b) shows $\mu$ in the transport regime, varying between $0\leq\mu\lesssim 0.5$ which increases for larger values of $\alpha_1$ and smaller $\vaf$. The contours indicate that $\mu$ remains constant on manifolds in $\alpha_1-\vaf$ space with positive slopes. Therefore with increasing $\vaf$, larger activity $\alpha_1$ are needed to achieve the same transport coefficient. Finally, we note that since the transport condition is an inequality, the transport coefficient $\mu$ undergoes a discontinuous transition as we cross the boundary manifold $(\alpha_1^*,\tauk^*,g_0^*,\vaf^*)$, except at the trivial boundary at the $(\vaf, g_0)\to(0,0)$ cross section.

\begin{figure}
    \centering
        \centerline{\includegraphics[width=8cm]{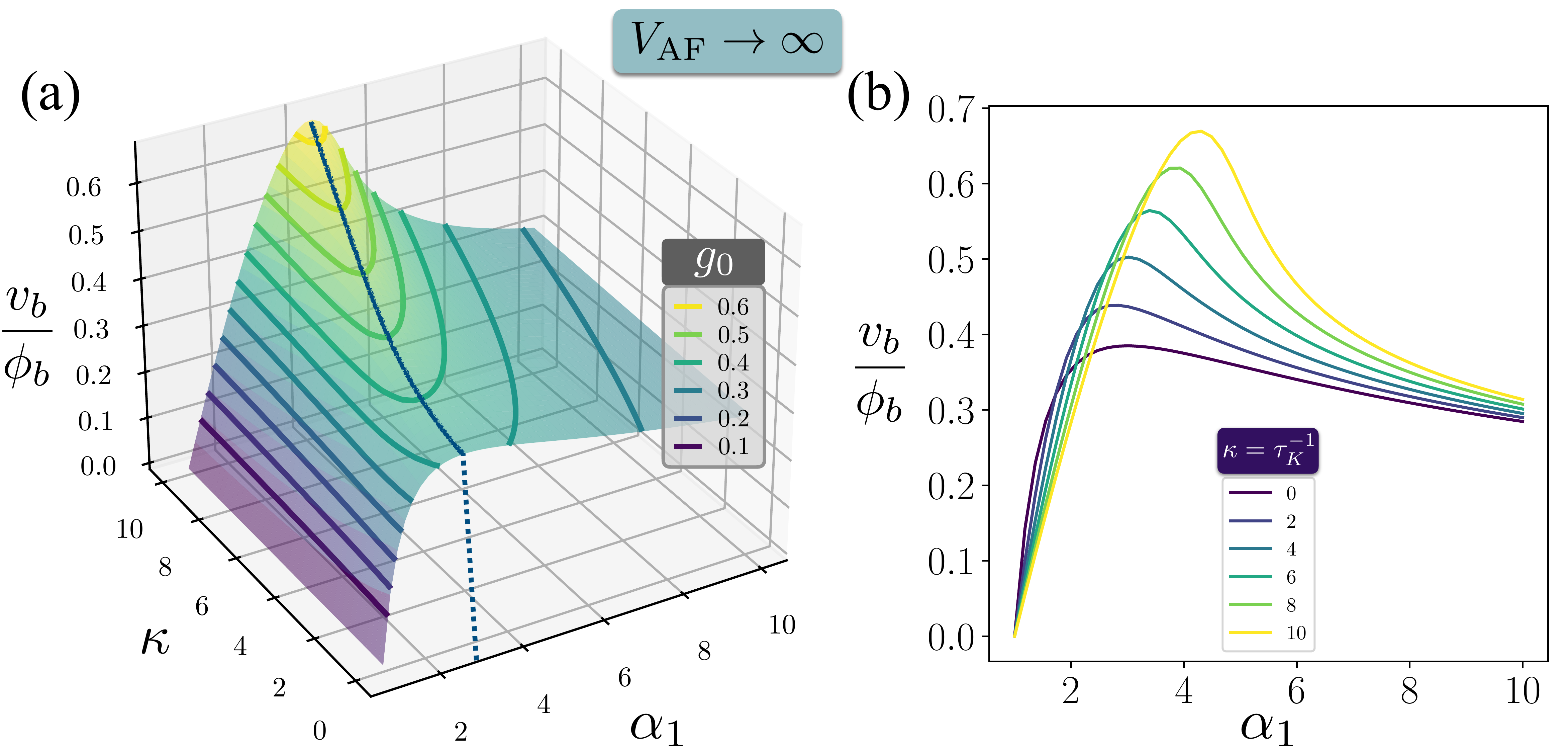}}
            \caption{Limit of $\vaf\to\infty$. (a) shows the function on the r.h.s. of Eq. (\ref{mt.eq.vafinftycond}) in terms of $(\alpha_1,\kappa)$. The contours show constant $g_0/\Gamma$. Panel (b) shows the cross sections of the surface at discrete values of $\kappa = \tauk^{-1}$, the inverse of the Kelvin timescale.} 
        \label{mt.fig.vafinftycond}
    \vspace{-3mm}
\end{figure}

\subsubsection{Limit of Large Activation Velocity}{\label{mt.sec.largeVaf}}
In this section, in order to explore the limit case of large $\vaf$, we adopt a self-consistent approach as follows: for given $(\alpha_1,\tauk,g_0)$, instead of finding the threshold velocity beyond which the transport fails, we try to solve the inverse problem by asking whether there exist regions in the phase diagram for which the threshold velocity diverges. Namely we assume that $\vaf\to\infty$, and see if we can find a solution to the parameters $(\alpha_1,\tauk,g_0)$ that satisfy the transport condition. In the limit of large $\vaf$, we can heuristically derive analytical solutions for $v_b$ and $\vdw$, the boundary and DW velocities, respectively. The assumptions and approximations are backed by numerical results; see the details in Appendix Sec. (\ref{sm.sec.infiniteVaf}). Using conservation of mass and momentum we derive the transport condition in terms of parameters $(\alpha_1,\tauk)$, and show that in the limit of $\vaf\to\infty$, the boundary and the DW velocities approach saturated values given by:
\begin{subequations}{\label{mt.eq.vbvdw}}
    \begin{gather}
        \lim_{\vaf\to\infty}v_b = \alpha_1^{1/2}\left(\frac{(\phi_0-\phi_i)(\phi_e-\phi_i)}{\phi_0\phi_e}\right)^{1/2}{\label{mt.eq.vb}}\\
        \lim_{\vaf\to\infty}\vdw = \alpha_1^{1/2}\left(\frac{\phi_e(\phi_0-\phi_i)}{\phi_0(\phi_e-\phi_i)}\right)^{1/2},{\label{mt.eq.vdw}}
    \end{gather}
\end{subequations}
where $\phi_0=\alpha_1/\alpha_2$ and $\phi_e$ is a function of $(\alpha_1,\tauk)$. From Eqs. (\ref{mt.eq.vbvdw}), we understand that $v_b$ and $\vdw$ are increasing and decreasing functions of $\tauk$, respectively. The boundary velocity increases with increasing $\tauk$, since the activated material spends more time in the fluid state with nonzero strain rate. The velocity of DW, on the other hand, increases as the Kelvin timescale decreases, namely the transition to rigid state occurs faster; this is because for smaller $\tauk$ the equilibrium (boundary) density approaches the initial density, thus the DW of the density field moves faster.

The transport condition, $v_b/\phi_b = v_b/\phi_e \leq g_0/\Gamma$, is now obtained by substituting the above relation for $v_b$ from Eq. (\ref{mt.eq.vb}); which yields:
\begin{equation}{\label{mt.eq.vafinftycond}}
    \frac{\alpha_1^{1/2}}{\phi_e}\left(\frac{(\phi_0-\phi_i)(\phi_e-\phi_i)}{\phi_0\phi_e}\right)^{1/2}   \leq \frac{g_0}{\Gamma}.
\end{equation}
The left hand side of the above inequality, which equals $\lim_{\vaf\to\infty}v_b/\phi_b$, involves $\alpha_1$ and $\phi_e$, thus is a function of $\alpha_1$ and $\kappa=1/\tauk$. Should $(\alpha_1,\tauk,g_0)$ satisfy the above relation in Eq. (\ref{mt.eq.vafinftycond}), the threshold velocity $\vaf^*$ diverges; and for triples $(\alpha_1,\tauk,g_0)$ that do not respect the inequality the threshold $\vaf^*$ remains finite, corresponding to the boundary surfaces between Transport and No-Transport sectors in Fig. (\ref{mt.fig.PD}). In Fig. (\ref{mt.fig.vafinftycond}), the l.h.s. is plotted against $(\alpha_1,\kappa)$. The contours show different values of $g_0/\Gamma$ equal to $g_0$ for $\Gamma=1$. As stated above, the maximum boundary velocity is obtained for $\kappa=0$, corresponding to purely viscous material:
\begin{equation}
    \frac{\alpha_1^{1/2}}{\phi_0}\left(1-\frac{\phi_i}{\phi_0}\right) \leq \frac{g_0}{\Gamma}.
\end{equation}
Setting $\phi_i=1$ and $\alpha_2=1$ and $\Gamma=1$, the relation reduces to: $(\alpha_1-1)/\alpha_1^{3/2}\leq g_0$. The expression on the l.h.s. is maximized at $\alpha_1=3$; coinciding with the position of the depression on the phase boundary manifold, in the left panel of Fig. (\ref{mt.fig.PD}a).  This non-monotonic behavior of the ratio $v_b/\phi_b$, at fixed values of $\tauk$, originates from the competition between $v_b$ and $\phi_b$, both of which are increasing functions of $\alpha_1$. At small values of $\alpha_1$, the increase of $v_b$ dominates that of $\phi_b$. At larger values of $\alpha_1$ (e.g. $\alpha_1 \to 3$ for $\tauk\to\infty$), the boundary density grows faster and dominates velocity increase; hence the ratio decreases. 

The inequality in Eq. (\ref{mt.eq.vafinftycond}) suggests that for a given magnitude of the field-particle coupling $g_0$, there exist regions of the phase diagram where the threshold velocity diverges; namely the coupling is strong enough that the active material---with the parameters within those regions---can transport the particle regardless of how fast the AF moves. The relation between $(\alpha_1,\tauk)$ or equivalently $(\alpha_1,\kappa)$, for which the transport condition is met is obtained by noting that the expression on the l.h.s. of Eq. (\ref{mt.eq.vafinftycond}) is maximized along a ridge; the dark blue curve in Fig. (\ref{mt.fig.vafinftycond}a). At a fixed value of $g_0$ (the contours in the same Figure), the threshold velocity diverges for the regions of the manifold that fall below $g_0/\Gamma$. This conditions is evidently achieved more easily for lower $\kappa$, in agreement with the phase diagrams shown in Fig. (\ref{mt.fig.PD}a), where we observe that smaller rigidity makes the transport more accessible. 

Lastly, we can understand how the velocity of the DW, $\vdw$, can tell us about the possibility of transport. For small $\tauk$ the equilibrium density in state 2 is closer to the initial density, namely the density difference between the states 1 and 2 is small, thus the DW travels faster (see Fig. (\ref{mt.fig.motion}a) for the definitions of states 1 and 2). This can also be inferred from Eq. (\ref{mt.eq.vdw}). On the other hand, since $\dot{\bep}_2 = 0$---at all times---the faster the DW moves, the longer the extension of the equilibrated region, the shorter the region with nonzero strain rate, and the smaller the boundary velocity $v_b$ (since $v_b$ is obtained by integrating over the strain rate $\int_{\Omega(t)}\dd x\,\dot\bep(x,t)$ along the extension of the active region). Therefore, larger $\vdw$ implies both smaller $v_b$ and $\phi_b$. At first glance it is unclear which of the two decrease faster with increasing $\vdw$. In fact, like in the case of $\alpha_1$, the ratio $v_b/\phi_b$ is non-monotonic with respect to $\tauk$. In Fig. (\ref{mt.fig.vafinftycond}b) we see that at small values of $\alpha_1$, the smaller the $\tauk$, the smaller boundary velocity dominates the ratio, making it more accessible; whereas for larger activities, small values of $\tauk$ make the boundary density so small that the active force on the particle cannot pull the particle against the viscous background. Therefore, the velocity of the DW signifies feasibility or failure of the transport, for the opposite reasons at small and large values of the activity. Finally, we shall add that in the limit of $\vaf\to\infty$, we have $\mu=V_p/\vaf\to 0$, namely the transport coefficient vanishes.

\section{Energetics and the First Law}
The analysis of the energy conservation in the transport problem requires careful considerations since the effect of the field-particle interaction is neglected on the field. Here we focus on the flow of different energy contributions in an isolated active field, which is activated all at once, and is undergoing symmetric contraction. We define different contributions for the kinetic and (passive) elastic energies, active and passive works, and dissipated energy; and demonstrate the conservation of energy. The First Law governing the different contributions to energies can be written in terms of the corresponding (inexact) differentials:
\begin{align}{\label{mt.eq.1stlaw}}
    \dd E + \left[\db W_a + \db W_p\right] + \db Q = 0.
\end{align}
Here $\db Q$ is the dissipated energy, and $\dd E$ (exact differential) is the change in the internal energy (sum of the kinetic and elastic energies). The terms $\db W_{a,p}$ correspond to the works performed by the active and passive stresses on the environment, respectively. The dissipated energy can also be calculated either directly by integrating over the viscous and drag terms, or indirectly using the above equation.

Each term in Eq. (\ref{mt.eq.1stlaw}) is obtained by integrating the corresponding density function over the spatial extension of the active region $\Omega(t)$; for a general functional: $\db \mathcal{H} = \int_{\Omega(t)} \dd x\, \db \mrm{h}$. The differential is found by $\db h = p_h \,\dd t$, where $p_h$ is the power density obtained by $p_h = \mathbf{f}_h.\bf{v}$, with $\mathbf f_h$ and $\mathbf v$, the corresponding force and velocity, respectively. The forces appear as different terms in Eqs. (\ref{mt.eq.constitutive}) \& (\ref{mt.eq.stress}). For the internal energy, active and passive works, and the dissipated energies we have:
\begin{subequations}
    \begin{gather}
        \mrm{e}=\frac{1}{2}\left(\phi_av_a^2 + \kappa\bep^2\right),\\
        \db \mrm{w}_{a,p}(x,t) = -v_a\nabla\cdot\sigma_{a,p} \dd t,\\
        \db \mrm{q}(x,t) = \eta\, v_a\nabla^2 v_a\dd t + \gamma v_a^2\dd t{\label{mt.eq.dQ}}.
    \end{gather}
\end{subequations}
where $\sigma_a=-\alpha_1\phi_a$ and $\sigma_p = +\alpha_2\phi_a^2$, are the active and passive stresses, respectively. In the above equation we introduce a {direct} relationship between the dissipation and the dissipative forces in Eq. (\ref{mt.eq.dQ}). As mentioned previously, another way to calculate the dissipated energy is through Eq. (\ref{mt.eq.1stlaw}). In Appendix Sec. (\ref{sm.sec.energetics}), we show that the two methods agree. Individual contributions to the total energy are shown in Appendix Fig. (\ref{sm.fig.energetics}).

\section{Discussion and Outlook}{\label{mt.sec.discuss}}
The limits of power extraction in our system is set by the properties of the active material, as well as the coupling to the particle $g_0$ and the particle's drag coefficient $\Gamma$. The material's properties include activity $\alpha_1$, the coefficient of the excluded volume pressure $\alpha_2$, and the viscoelastic timescale $\tauk=\kappa^{-1}$. The effect of the latter, as shown in Fig. (\ref{mt.fig.PD}b), indicates that for fixed $(\alpha_1,g_0)$, decreasing the timescale lowers the maximum velocity $\vaf^*$. This result seems counter-intuitive, as solids, generically, tend to transmit forces over larger length scales. In our system, however, the timescale (inverse of the rigidity) also influences the equilibrium density which in turn determines the force experienced by the particle at the boundary. For $\tauk\to 0$, we have $\phi_e\to\phi_i$, which results in smallest force on the particle. On the other hand, for fixed $(\tauk,g_0)$, the transport coefficient $\mu$ increases with $\alpha$. In other words, even though the threshold velocity decreases because of small boundary density, the ratio increases in the transport regions. A very important point is that the steady-state ratio $v_b/\phi_b$ is only the determinant of transport condition, in the specific scenario where $g_0>0$, for which the particle is trapped at the boundary. For negative coupling $g_0<0$, the condition reduces to $v(X_p)/\nabla\phi(X_p)\leq g_0\ell/\Gamma$, where the velocity and density gradient are evaluated at the particle's position.

As discussed above, what determines the boundary between transport and no-transport regions is the ratio $v_b/\phi_b$. While $\phi_b=\phi_e$ is only a function of the material properties $(\tauk,\alpha_1,\alpha_2)$, the boundary velocity $v_b$ depends on $\vaf$ as well; rendering it a complicated and nontrivial variable. The boundary velocity increases monotonically with $\vaf$ and was shown to saturate at a constant value which is independent of $\vaf$. The transport fails for $v_b>v_b^*=\phi_eg_0/\Gamma$, which corresponds to a threshold AF velocity $\vaf^*$. In this paper we addressed two distinct problems: (1) possibility of transport, and (2) transport coefficient. The two differ in that, in the first case, the only condition is to make the ratio $v_b/\phi_b$ as small as possible, where as (2) requires $v_b$ to grow as larger as possible while respecting the inequality. 

Since the rate of the injected energy is proportional to $\vaf$, the existence of the \textit{upper} bound (as opposed to lower bound) on $\vaf$ seems counterintuitive; as it implies that increasing the rate of injected energy above a threshold results in the failure of transport. This is because the active region is an open system, the mass and spatial extension of which are constantly increasing. Therefore, for larger $\vaf$ the injected energy requires to travel over progressively larger and larger distances. Another interesting observation is that the coupling $g_0$ only determines the boundaries of the transport sector, not the transport coefficients. This is because once $g_0$ surpasses the transport threshold, the particle is trapped and transported with the boundary velocity; see the contours of constant transport coefficient in Fig. (\ref{mt.fig.PD}b).

The phenomenological parameters of our model can be qualitatively traced back to the system-specific microscopic mechanisms. For example, in a cross-linked active network, the magnitude of the active force dipoles depend on the walking velocity of the motors; and the Kelvin viscoelastic timescale depends on the rate of gelation and the assembly of the network, namely how quickly the cross-linking takes place and the elastic response dominates the viscosity, which itself depends on the concentration of cross-linkers, the chemical bonding energy, and the ambient temperature among other factors. Therefore, in the case of active networks, the longer the rigidification takes, the faster the transport of the particle. Inferring the microscopic parameters requires a more accurate map between the microscopic and the coarse-grained parameters, which warrants separate numerical and experimental works. Our framework provides a broader picture which can be interpreted in different situations; therefore, it helps with searching for system-specific ranges of parameters to achieve optimal efficiency and/or transport coefficients.

Our model comes with some limitations rooting from the approximations we have made. Most importantly we have assumed (i) zero drag for the field, $\gamma\to 0$, and (ii) ignored the effect of particle on the field. We can intuitively speculate how results would change if (i) and (ii) are lifted. For fixed parameters, both forces would impede the velocity $v(x)$ everywhere along the active region including the boundary, counteracting contraction. Therefore, $\vaf^*$ increases, and the transport phase occupies larger subspace of the phase diagram; it takes larger $\vaf$ for the particle to escape the potential. Furthermore, we neglected the effect of thermal fluctuations, which facilitates escape of the particle from the potential well. Speculatively, as the temperature rises the transport region of the phase diagram shrinks towards smaller $\vaf$ and larger activity. 

An interesting question is the maximum extractable energy from the \emph{bulk} of the active region. Should the coupling to the bulk of an active system be feasible, the extracted energy increases, possibly scaling with the system's volume. By definition, however, coupling to the bulk requires extended domains, and thus the extracted energy would be delocalized.

\textit{Perspective and Outlook.---}Active systems adopt a diverse set of structural and mechanical phases, that can, in principle, be harnessed for the performance of tasks including mechanical actuation, transport, and mixing \cite{wu2017transition,needleman2017active,ross2019controlling,liu2024force}. Extensile active systems such as active nematics, internally generate fluid flows useful for material transport and mixing \cite{wu2017transition,qu2021persistent,lemma2023spatio}. Recent experiments demonstrate the controllability of active systems using external sources such as light. While optimal control strategies have been explored theoretically in specific situations (e.g., active nematics, Brownian particles), fundamental limits on active system performance, including macroscopic work extraction given microscopic energy consumption, remain less understood \cite{norton2020optimal,shankar2022optimal, giomi2014spontaneous,omar2018swimming}. This study establishes energy extraction bounds in contractile active matter. Extending these bounds across phases and tasks is vital to determine general limits in terms of abstract quantities: the active system's structure, microscopic interactions, and information availability about the system state that can be acquired by an external observer or control system \cite{schildknecht2022reinforcement}.

Our results extend beyond physics and engineering, offering insights into the design principles of biological systems. Cells utilize active materials like motor-filament networks for tasks such as material transport, cell division, and shape changes \cite{needleman2017active}. Optimality principles, a key lens for biological mechanisms, guide this exploration \cite{bialek2017perspectives}. Biological active materials contain a wide variety of protein components, including active cross-linkers and filaments which influence viscosity and elasticity. As we demonstrate that limits on energy extraction depend on the microscopic composition of an active system, it will be interesting to ask whether biological systems have evolved components or regulatory strategies that maximize the efficiency of energy extraction from internal active systems or instead, compromise energy efficiency for other goals including the accuracy of internal processes and robustness to molecular fluctuations \cite{petkova2019optimal,hopfield1974kinetic}. Our study is intended to provide a framework for quantitative investigation of such questions. 

\section*{Acknowledgments}
This work was supported by Packard Foundation, Rosen Center for Bioengineering, Moore Foundation, and Heritage Medical Research Institute. We would like to thank Foundational Questions Institute and Fetzer Franklin Fund through FQXi 1816 for funding the research.

\PRLsep
\appendix
\renewcommand\thefigure{\thesection.\arabic{figure}}    
\counterwithin{figure}{section}

\section*{Appendix}
In the Appendix, we first discuss the model proposed for the active model and demonstrate some analytical and numerical results. As explained in the Main Text (MT), we focus on active filament networks treated commonly as viscoelastic substance, and suitable for describing cytoskeletal materials. In particular we choose Kelvin-Voigt viscoelasiticity, to capture the early fluid-like response of the assembling network. For future reference, we tabulate the parameters and variables in Table (\ref{sm.table1}).

\section{Kelvin-Voigt Model of Viscoelasticity}{\label{sm.sec.model}}
In this section we introduce the Kelvin-Voigt model which captures the creep behavior in viscoelastic materials, such that it shows fluid-like viscous response on short timescales over $\tau_K$, and restoring elasticity on large timescales. Consider the constitutive equation for Kelvin-Voigt model with strain and stress fields denoted by $\bm{\mathcal{E}}$ and $\bsg{}$ reads:
\begin{equation}
    \bsg{} = \eta\left(\frac{\bm{\mathcal{E}}}{\tau_K} + \frac{\text{D}\bm{\mathcal{E}}}{\text{D}t}\right).
\end{equation}
With $u(x,t)$ the displacement field that satisfies $\bm{\mathcal{E}} = \nabla \bm u$, the conservation of momentum reads:
\begin{equation}
    \DD(\phi_a\vv)=\nabla\cdot\bsg{} + \ff,
\end{equation}
where $\ff$ denotes the body force, which his obtained by:
\begin{equation}
    \ff=\nabla\cdot\bsg{int}-\gamma_0\phi_a\phi_p(v_a-v_p).    
\end{equation}
\begin{figure*}[ht]
    \centering
        \centerline{\includegraphics[width=18cm]{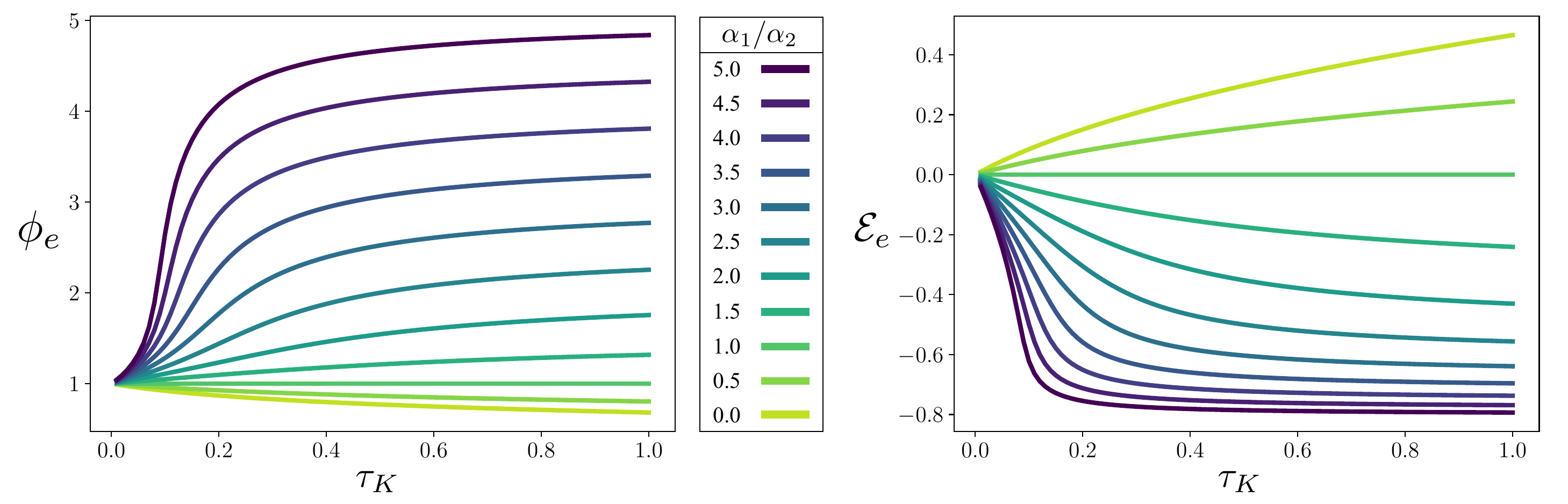}}
            \caption{The equilibrium densities (left) and strains (right) plotted against Kelvin time $\tauk$, for different values of $\alpha_1/\alpha_2 = \phi_0$.}
        \label{sm.fig.straineq}
    \vspace{-3mm}
\end{figure*}
The first term is the divergence of the internal stress which is introduced below. The second term is the drag force between the active and passive components of densities $\phi_{a,p}$. The functional form of the drag force is chosen to respect symmetries: odd under $v_a \leftrightarrow v_p$, and even under $\phi_a \leftrightarrow \phi_p$. The total density equals $\Phi = \phi_a + \phi_p$. We also note that the two components moving against each other respect mass conservation: $\phi_av_a + \phi_pv_p = 0$. Therefore we get:
\begin{align}{\label{sm.eq.drag}}
    f_{\text{drag}} &= -\gamma_0\phi_a\phi_p(v_a-v_p)\nonumber\\
    &= -\gamma_0\phi_a(\phi_pv_a - \phi_pv_p)\nonumber\\
    & = -\gamma_0\phi_a(\phi_pv_a + \phi_av_a)\nonumber\\
    & = -\gamma_0\phi_av_a(\phi_p + \phi_a)\nonumber\\
    & = -\gamma_0\Phi\,\phi_av_a = -\gamma \phi_a v_a.
\end{align}
The conservation of mass is also expressed by continuity equation:
\begin{equation}
    \DD \phi_a + \phi_a\nabla\cdot \vv = 0.
\end{equation}
Finally we introduce the internal stress as an expansion in the active density:
\begin{align}{\label{sm.eq.virial}}
    \bsg{int} = \sigma^{\text{int}}(\phi_a) \, \mathbb{I} &= -\alpha_1 \phi^{\phantom{}}_a\mathbb{I} + \alpha_2 \phi_a^2\mathbb{I} + \mathcal{O}(\phi_a^3)\\
    &= -\alpha_1\phi_a(1-{\phi_a}/{\phi_0})+ \mathcal{O}(\phi_a^3),
\end{align}
where $\phi_0 = \alpha_1/\alpha_2$. Hereafter, we reduce the tensors to scalar in 1D, thus $\sigma$ and $\varepsilon$ represent $\bsg{}$ and $\bm{\mathcal{E}}$, respectively. We can solve the dynamics of a 1D active string of length $2L_i$ and initial density $\phi_i$, which is governed by the continuity and constitutive equations. Using the former we have $\phi_iL_i=\phi_a L=m$; and definition of $\bep=L/L_i - 1$. Thus the relation between density and strain can be written as: $\phi_a = \frac{\displaystyle\phi_i}{1+\bep}$. In equilibrium we have $\DD\bep_e = 0$. Therefore have: 
\begin{equation}
    \frac{\eta}{\tauk}\,\bep_{e} = - \alpha_1\left(\frac{\phi_i}{1+\bep_{e}}\right) + \alpha_2\left(\frac{\phi_i}{1+\bep_{e}}\right)^2.
\end{equation}
The above equation admits $\bep_{{e}}=0$ for $\phi_0=\phi_i$. If $\phi_0>\phi_i$, the contractile stress dominates the extensile stress, and the equilibrium strain is negative. In the following we are going to work in this regime where $\phi_0\geq\phi_i$. 
Alternatively in terms of the density field we find in equilibrium: 
\begin{equation}
    \phi_e^3 - \phi_0\,\phi_e^2 + \left(\frac{\eta\phi_0}{\tauk\alpha_1}\right)\phi_e - \left(\frac{\eta\phi_0\phi_i}{\tauk\alpha_1}\right) = 0.
\end{equation}
The solutions to this equation are plotted in Fig. (I.1). In the case of $\alpha_1/\alpha_2=\phi_0=\phi_i$ the equilibrium density $\phi_e=\phi_i$; which corresponds to the case of neither contraction nor extension. In the limit of $\tauk\to\infty$, corresponding to fully viscous system, we obtain $\phi_e\to\phi_0$. In the opposite limit $\tauk\to0$, we get $\phi_e\to\phi_i$. The latter represents an infinitely rigid system, where the entire energy of stress is stored in a infinitesimal strain. Therefore the equilibrium changes from $\phi_e=\phi_i$ to $\phi_0$ as $\tauk$ varies between $0$ and $\infty$.

Using the relation between the strain and density: $\phi\,(1+\bep)=\phi_i$, we can find the equilibrium density, satisfying: $\phi_e^3 - \phi_0\,\phi_e^2 + \left(\frac{\eta\phi_0}{\tauk\alpha_1}\right)\phi_e - \left(\frac{\eta\phi_0\phi_i}{\tauk\alpha_1}\right) = 0$. We define the dimensionless parameter $\nu={\alpha_1\tauk\phi_0}/{\eta} = \alpha_1\phi_0/\kappa$. In the absence of the elastic contribution $\kappa\to0$, i.e. $\nu\to\infty$ the asymptotic solution reads: $\phi_e = \phi_0 - (\phi_0-\phi_i)\,\nu^{-1} + \mathcal O(\nu^{-2})$. Therefore $\phi_e \to \phi_0=\alpha_1/\alpha_2$. In the opposite limit $\nu\to 0$, we get: $\phi_e=\phi_i+(\phi_i^2/\phi_0)\left(1-{\phi_i}/{\phi_0}\right)\nu+\mathcal O(\nu^{2})$. Thus for large rigidity, $\phi_e \to \phi_i$ the equilibrium density remains close to the initial density. We will see below that this becomes important when interpreting the results. To summarize, we have:
\begin{gather}{\label{sm.eq.densityexp}}
    \phi_e = \phi_0 - (\phi_0-\phi_i)\nu^{-1},\nonumber\\
    \phi_e = \phi_i + (\phi_i^2/\phi_0) (1-\phi_i/\phi_0)\nu.
\end{gather}

\subsection{Dimensional Analysis}
We can simplify our equations by finding the dimensionless parameters. To this end we rescale $x,t$ and $\phi$ as follows: $x = x_s\tilde x$, and $t = t_s\tilde t$ and $\phi = \phi_s\tilde\phi$; all variables e.g. $f$ are rescaled accordingly and denoted by $f = f_s\tilde f$. The scaling factor of density is chosen to be equal to $\phi_s = \phi_i$; i.e. the initial density.
\begin{align}
    \frac{x_s\phi_i}{t_s^2}\text{D}_{\tilde t}(\tilde \phi \tilde{\dot v}) + \frac{\eta}{\tauk x_s} {\partial}^2_{\tilde{x}} \tilde u + \frac{\eta}{t_sx_s}{\partial}^2_{\tilde{x}}\tilde v -\frac{\gamma x_s}{t_s}\tilde v& \nonumber\\
    = -\frac{\alpha_1\phi_i}{x_s}\left[1 - 2(\alpha_2/\alpha_1){\phi_i}\tilde\phi\right]{\partial}_{\tilde x}\tilde{\phi}&\;.
\end{align}
The parameters of interest, that span the phase diagram are activity $\alpha_1$ and the Kelvin timescale $\tauk$. As such we obtain the scaling factors:
\begin{align}
    x_s = \frac{\eta}{\phi_i\alpha_2^{1/2}}\qquad \text{and} \qquad t_s = \frac{\eta}{\phi_i^2\alpha_2}.    
\end{align}
The characteristic velocity is thus given by $v_s = \phi_i\alpha_2^{1/2}$. The above equation can be recast in the following form (tildes are dropped hereafter):
\begin{equation}
     \partial_x^2 v + \varphi\DD (\phi \dot v) + \kappa \,\partial_x^2 u - \gamma v = -\alpha (1-2\phi/\phi_0)\,\partial_x \phi,
\end{equation}
where
\begin{align}
    &\varphi= \phi_i,\qquad\hspace{1.05cm}\kappa = \left(\frac{\eta}{\phi_i^3\alpha_2}\frac{1}{\tauk}\right),\qquad\hspace{0cm}\nonumber\\&\gamma= \left(\frac{\eta\gamma_0}{\alpha_2\phi_i^3}\right),\qquad\hspace{0cm}\alpha = \left(\frac{\alpha_1}{\alpha_2\phi_i}\right).
\end{align}

The above parameters $\varphi$, $\kappa$, $\gamma$ and $\alpha_1$ are the four dimensionless parameters, which characterize the mass of the field, rigidity of the field, the drag coefficient and the activity, respectively. Furthermore, in order to specifically focus on the role of rigidity and its competition with viscous, and active terms, we assume that $\gamma\to0$.

A simplifying picture to gain intuition into the dynamics of system is provided in MT Fig. (\ref{mt.fig.motion}). Here we illustrate the transition of the density field as the newly activated material moves from initial state 1 to final state 2, separated by a rather sharp transition which is concomitant of the increase in the density field in time (in material coordinates), which is now projected in position space: namely different segments at each snapshot of the system can \emph{roughly} be viewed as a segment moving in time. We call the separation region the Domain Wall (DW). The velocity of the DW, by definition, lies in between the velocities of the Activation Front (AF) and that of the boundary, and is determined by field parameters, activity, and rigidity. Using a heuristic argument we derive the analytical solution to the boundary and DW velocities in the limit of large velocities of the AF.

\section{Field-Particle Interaction}{\label{sm.sec.ParticleInteraction}}
The particle is assumed to be embedded in and interacting with both active and passive components. Thus the motion of the particle in the mixture of active and passive solutions is resisted by two drag forces, which are proportional to the densities and relative velocities (Stokes' law), and a constant $\Gamma_0$. Denoting the particle's mass and velocity by $M$ and $\vp$, respectively, we get:
\begin{align}{\label{sm.eq.particle}}
    M\dot{\bf{V}}_p + \Gamma_0\phi_a(\vp-\vv_a) + \Gamma_0\phi_p(\vp-\vv_p) &= \ff_a + \ff_p.
\end{align}
The r.h.s. of the above equation $\ff_{a,p}$, are the two forces exerted on the particle as a result of interaction with active and passive components. We will derive them below. The first term on the l.h.s. captures inertia of the particle. The sum of the second and third terms on the l.h.s. can be simplified by noting that the conservation of mass demands: $\phi_a\vv_a + \phi_p\vv_p = 0$, and $\phi_a + \phi_p = \Phi$ is constant. Therefore, 
\begin{equation}
    \ff_{\text{drag}} = \Gamma_0\phi_a(\vp-\vv_a) + \Gamma_0\phi_p(\vp-\vv_p) = \Gamma_0\Phi\vp \equiv \Gamma\vp.
\end{equation}

The simplest form of the coupling to density is taken to be proportional to the density field on the particle's surface; namely ${\bm f}=G\phi(\rr)\nn$, where $G$ is the coupling constant at the particle's surface $\partial P$; and $\nn$ is the unit normal vector. 
\begin{equation}
    \ff_{a,p}=\int_{{\partial P}}\dd^2r\,\nn\, G_{a,p}\,\phi_{a,p}.
\end{equation}

Using Stokes' theorem the surface integral can be written in the form of a volume integral: $\int\dd VG\nabla\cdot\phi\nn$. For a particle of size $\ell$, much smaller than the length scale of the density gradient, $\ell\ll\phi(\RR)/\nabla\phi(\RR)$, we take $\nabla\phi$ to be constant over the boundary of the particle and the total force turns out $\ff=g_0\ell\nabla\phi$, where $g_0\propto G$. The sum of the passive and active contributions can be found by noting that $\nabla\phi_a = -\nabla\phi_p$. Therefore we get:
\begin{equation}
    \ff_a+\ff_p = g_a\ell\nabla\phi_a + g_p\ell\nabla\phi_p = (g_a-g_p)\nabla\phi_a = g_0\ell\nabla\phi_a.
\end{equation}
As mentioned in the MT, we focus on the case of $g_0>0$. Below we explain what happens in the case of $g<0$. Furthermore, when the particle is appended to the boundary of the active region, the gradient of the density is proportional to density itself $\nabla\phi(\RR)\propto\ell^{-1}\,\phi(\RR)\int_{\partial P}\nn$ for $\ell$ the linear size of the particle. Thus the particle-field coupling {\emph{at the boundary}} reduces to $g_0\phi(\RR)\xx = g_0\phi_b\xx$, where $\xx$ is the average direction of the density gradient over the surface of the particle.

Using the above force field, we can define the corresponding potential energy to which the particle is subjected. Choosing the reference energy to be that of a free particle, the potential field is obtained by integrating the force: $\uu = -g_0\int^xdx\nabla\phi_a = -g_0\phi_a$. The potential is thus proportional to the negative of the density field and has a sharp jump at the boundary of the activate region, where the AF starts moving.

The density field generated by an AF moving in the $+\xx$ direction has a sharp gradient at its free boundary, resembling a step function, which is followed by negative gradient due to the asynchronous activation along $+\xx$ direction. Therefore, assuming that the initial position of the AF and the position of the particle are the same, the particle remains always trapped at the boundary and cannot move faster than the boundary. Namely, upon slightly faster motion, the particle is exposed to the negative density gradient which pushes the particle back to the boundary. If the activation process is carried out continuously ahead of the moving particle, its motion can potentially be sustained with constant velocity.

The above picture explains why we are interested in the regime of $g_0>0$. In the opposite limit of $g_0<0$ in which the particle is repelled by the active density field. The repulsive force pushes the particle away from the boundary, the particle would get detached, hence no transport.

The effective coupling to the particle which is parametrized by a constant $g_0$, is obtained by integrating the a pressure-like quantity---induced by the field---over the surface of the particle, which yields $\sim\nabla\phi_a$. The magnitude of such force is particularly dependent on the microscopic interactions between the active medium and the surface of the particle, is taken to be phenomenological, and cannot be expressed in terms of the internal interactions of the active medium alone. This is unlike the classical example of a Brownian particle embedded in a fluid where the coupling to the particle, and thus the diffusion constant, can be expressed in terms of the internal pressure of the fluid. The reason behind this discrepancy is that in our system, the internal stress (pressure) of the active medium is a result of microscopic active force dipoles that are induced by cross-linkers walking along the filaments. The underlying mechanism of the interaction with the particle, however, need not be of the same origin. On the contrary the source of pressure experienced by a Brownian particle is the same as that experienced by the fluid's constituents themselves.

We would also like to mention here that our theory was originally inspired by experiments (see Ref. [14] in MT), showing the transport of microtubule asters. Asters are dense structures formed as the result of the collapse of an active network of polar filaments, under the contractile active stress generated by the active cross-linkers. In these experiments an initially localized aster is dragged along a trajectory using continual activation of the surrounding medium, exactly like what we described in our model. In this case, the microtubule aster can be treated like a separate particle, the internal modes of which are negligible and we only care of translational mode. As such the aster plays the role of the particle in our model. In such a system, the ``particle'' and the field are made of the same substance and therefore, the field-particle coupling can be expressed in terms of the same parameters as for the field alone. Furthermore, the contractile nature of the active stress guarantees that the effective interaction $g_0$ to be positive.

\section{Transport Criteria}{\label{sm.sec.transport}}
\begin{figure*}
    \centering
        \centerline{\includegraphics[width=17cm]{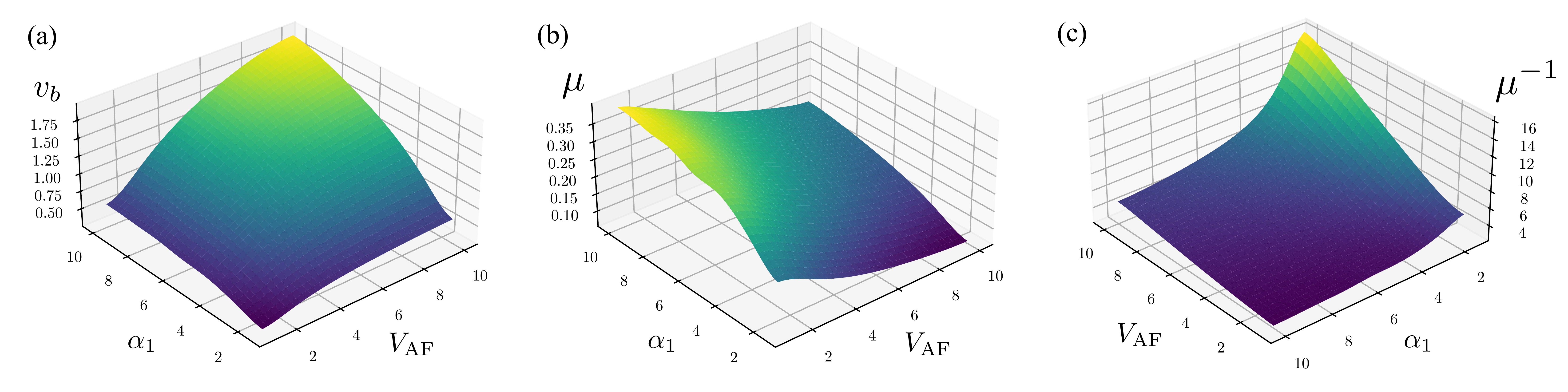}}
            \caption{For $\tauk\to\infty$ and $\gamma=0$, the surface plots illustrate (a) $v_b$, (b) $\mu$, and (c) $\mu^{-1}$, as functions of $\alpha_1$ and $\vaf$ (note the difference in labels and directions of $\alpha_1,\vaf$ axes in (c)). The factor $\mu^{-1}$ appears in the expression derived for $\vaf^*$ in MT Eq. (3).}
        \label{sm.fig.mu}
    \vspace{-3mm}
\end{figure*}
Transport of the particle demands that the force exerted on the particle at the boundary be greater than, or equal, to the drag force. If the boundary force is greater than the drag force the particle moves at velocity $V_p>v_b$ which puts the particle in a position with negative density gradient. The negative gradient pushes the particle back towards the boundary; making the boundary a stable minimum in the frame of reference comoving with the boundary velocity. The above condition leads to the relation we introduced in the MT: $v_b/\phi_b\leq g_0/\Gamma$.

As we explained in the Main Text, the boundary density $\phi_b(t;\alpha_1,\vaf)$ and velocity $v_b(t;\alpha_1,\vaf)$ are functions of parameters $\alpha_1,\vaf$ besides $\tauk,\gamma$. The long-time limits read $\displaystyle\lim_{t\to\infty}(\phi_b)=\phi_e$, and $\displaystyle\lim_{t\to\infty}(v_b)=\mu(\alpha_1,\vaf)\vaf$, where $\mu(\alpha_1,\vaf)<1$ is a dimensionless parameter that depends on activity and velocity of the AF. The dependence of $\mu$ is show in Fig. (\ref{sm.fig.mu}).

We can recast the transport condition in terms of $\mu$; while keeping in mind that $\mu$ is not a constant and itself depends on the parameters of the problem. In the limit of zero rigidity, the boundary density equals $\phi_b = \phi_0 = \alpha_1/\alpha_2$. In the opposite limit, we get $\phi_b = \phi_i$.

\begin{figure*}[ht]
    \centering
        \centerline{\includegraphics[width=17cm]{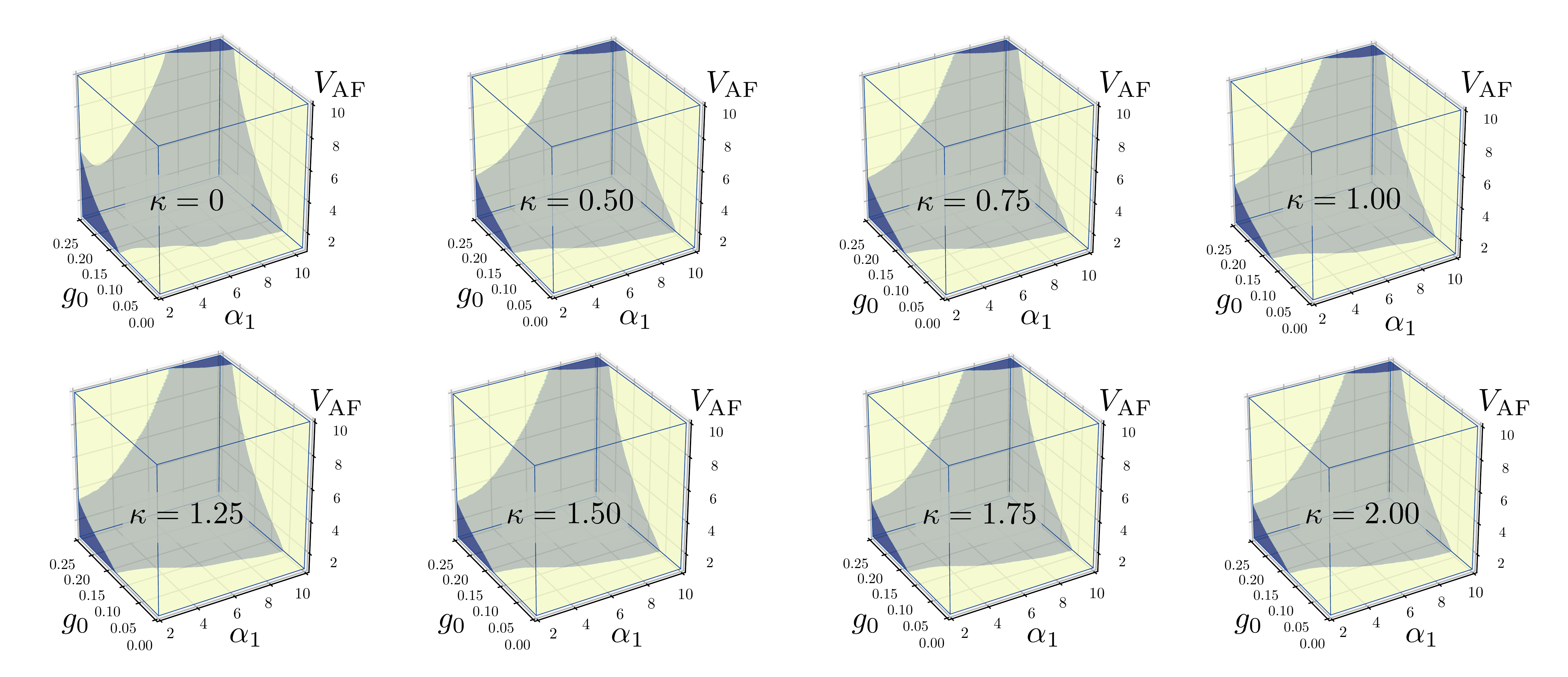}}
            \caption{Phase diagram for $\kappa=0,0.5,0.75,1,1.25,1.5,1.75,2$. The surfaces seem to be similar to a large extent, with minimal discrepancies. Increasing $\kappa$ is equivalent to decreasing $\tauk$ which varies here between $\tauk=0.5$ and $\tauk\to\infty$.}
        \label{sm.fig.kk_series}
    \vspace{-3mm}
\end{figure*}

\subsection{Analytical Solution}{\label{sm.sec.infiniteVaf}}
In this section we derive solutions to the velocities of the boundary $v_b$, and the DW, $\vdw$ in the limit of large $\vaf$. First, we introduce the following labels, for a right-moving AF. The domain of the system is divided into two segments, separated by the DW. The left segment is defined as $x_{\ell} \in [x_b,\xdw)$, and the right segment is defined as $x_r \in (\xdw,X_{\text{AF}}]$; where we used subscripts $l,r$ for the two segments, hereafter. Note that since the position of the DW is not a single point, but has finite width. For small widths of DW is, however, unimportant in our calculation. The left segment is defined as the region where the system has reached equilibrium where $\phi_{\ell}=\phi_e$, $\bep_{\ell}=\bep_e$ and $\dot\bep_{\ell} = 0$. Furthermore, since the left segment has reached equilibrium, its internal dynamics are absent, thus it can be treated as a rigid body moving with stead-state velocity $V_{\ell}$, which is essentially the same as the boundary velocity. The density, strain rate and the velocity of the right segment $(\phi_r,\dot\bep_r,V_r)$ are not trivial and depend on the parameters $(\alpha_1,\alpha_2,\tauk,\vaf)$. The reason for deviation of $\phi_r,V_r$ from $\phi_i,0$ is that the right and left segments are separated by a stress gradient at the DW and thus pull on each other. In actuality, the density of the right segment $\phi_r$ depends on position $x$. For sharp DWs, however, we can assume it is uniform in space. For large $\vaf$, which leads to large $V_{\ell},\vdw$, we can assume that $\phi_r = \phi_i$ and $\dot\bep_r=0$ and $V_r = 0$. This is intuitively because the force from the left on the right segment does not get enough time to contract and move the right segment; these assumptions are backed by the numerical results. Therefore, we have $(\phi_{\ell},\dot\bep_{\ell}) = (\phi_e,0)$, and $(\phi_r,\dot\bep_r,V_r) = (\phi_i,0,0)$. The unknowns are $V_{\ell}$ and $\vdw$; which can be found using the mass conservation. Denoting the lengths of the left/right segments by $L_{l,r}$, we have the following equation:
\begin{equation}{\label{sm.eq.masscons}}
    \phi_{\ell}\dot L_{\ell} + \phi_r\dot L_r = \phi_i \vaf.
\end{equation}
Furthermore, we have $\dot L_{\ell} = \vdw - V_{\ell}$ and $\dot L_r = \vaf - \vdw$. Substituting $\dot L_{l,r}$ in Eq. (\ref{sm.eq.masscons}), we get:
\begin{equation}
    \vdw = \left(\frac{\phi_e}{\phi_e-\phi_i}\right)V_{\ell}.
\end{equation}
On the other hand, and using the fact that the steady-state velocity of the left segment remains constant in time, we apply the Newton's second law to the left segment:
\begin{equation}
    \dot P_\ell = -\hat x\cdot \bm \Sigma[\phi_r] = \sigma[\phi_i] = \alpha_1\phi_i(1-\phi_i/\phi_0),
\end{equation}
where $P_\ell=\phi_\ell L_\ell \vl = \phi_e L_\ell \vl$ denotes the momentum of the left segment. Therefore we have:
\begin{align}
    \phi_e \vl \dot L_\ell = \phi_e \vl (\vdw-\vl) = \alpha_1\phi_i(1-\phi_i/\phi_0).
\end{align}
Using the relation between $\vdw$ and $\vl$ found above, we find:
\begin{equation}
    \vl = \alpha_1^{1/2}\left(\frac{(\phi_0-\phi_i)(\phi_e-\phi_i)}{\phi_0\phi_e}\right)^{1/2}.
\end{equation}
and the DW's velocity:
\begin{equation}
    \vdw = \alpha_1^{1/2}\left(\frac{\phi_e(\phi_0-\phi_i)}{\phi_0(\phi_e-\phi_i)}\right)^{1/2}.
\end{equation}
One important observation is that the zeroth order approximation of $\vl$ and $\vdw$, in terms of $\vaf^{-1}$, do not vanish; namely $\vl$ and $\vdw$ reach constant values. Now, we note that in the limit of $\tauk\to\infty$, we have $\phi_e \to \phi_0$, thus we get:
\begin{gather}
    \vl = \alpha_1^{1/2}\left(1-{\phi_i}/{\phi_0}\right)\nonumber\\
    \vdw = \alpha_1^{1/2}. 
\end{gather}
In the opposite limit $\tauk\to 0$, we have $\phi_e\to 0$, so we get $\vl\to 0$ to zeroth order in $\tauk$; which leads to $\vl\to 0$ and $\vdw\to\infty\lesssim\vaf$. We can also further investigate these expressions in powers of $\tauk^{-1}$ and $\tauk$, respectively, by using the equilibrium density we found in Eqs. (\ref{sm.eq.densityexp}). Since $\vl$ and $\vdw$ are monotonically increasing functions of $\vaf$, the maximum transport velocity over both $\vaf$ and $\tauk$ is obtained by
\begin{equation}
    \max_{\tauk}\vl = \alpha_1^{1/2}\left(1-{\phi_i}/{\phi_0}\right).
\end{equation}
Using the transport criteria: $\Gamma V_p = \Gamma \vl \leq g_0\phi_b = g_0\phi_\ell = g_0\phi_e = g_0\phi_0$, we conclude that the maximum $\vaf$ diverges provided that
\begin{equation}{\label{sm.eq.transcond}}
    \alpha_1^{1/2}\phi_0^{-1}\left(1-{\phi_i}/{\phi_0}\right) \leq g_0/\Gamma.
\end{equation}
From the above result we also obtain the maximum transport power:
\begin{equation}
    \max\pwr_t = \Gamma {V^*_p}^2 = \Gamma \alpha_1 (1-\phi_i/\phi^*_0)^2,
\end{equation}
where $\phi^*_0$ satisfies the equality condition in Eq. (\ref{sm.eq.transcond}).

\subsection{Effects of Non-Zero Drag and Finite Viscoelastic Timescale}{\label{sm.sec.drag}}
In the MT we demonstrated the phase diagram for $\kappa = 1$ (corresponding to $\tauk = 1$), and specifically focused on $\gamma=0$. Here we discuss how the phase diagram varies for multiple values of $\kappa$ and for $\gamma = 1$. In the MT we speculated how these parameters alter $\vaf^*$ for a given $g_0,\alpha_1$. 

\subsubsection{Non-Zero Drag of the Density Field}
In this subsection we demonstrate how nonzero drag force $\gamma \ne 0$ on the density field affects the results we obtained in the MT. As introduced in Sec. ({\ref{sm.sec.model}}), the simplest drag force that respects the symmetries under the exchange of active and passive components $\{a\}\leftrightarrow\{p\}$, is shown to be the following form: $f_{\text{drag}} = -\gamma\phi_av_a$. Moreover, in MT we speculated that at any given $\vaf$, nonzero drag force along the active region impedes the boundary velocity $v_b$, while $\phi_b$ would be left intact. Therefore, the $\mu$ function is a decreasing function of $\gamma$. This would increase $\vaf^*$. 

\subsubsection{Finite Kelvin Timescale}
Here we show how finite Kelvin timescale, i.e. $\tauk \not\rightarrow \infty$ and $\kappa \ne 0$, slightly alters the phase diagram and $\vaf^*$ in a predictable. Upon closer investigation, we see that by increasing $\kappa$, the transport region is pushed further towards smaller values of $\vaf$ (namely $\vaf^*$ decreases), and larger values of $\alpha_1,g_0$. The reason for this behavior is that $\phi_b=\phi_e$ decreases with increasing $\kappa$, as the rigidity stores some energy. Decreasing $\phi_b$ results implies that maximum $v_b$ decreases for the transport to be possible. This can also be seen by following the phase boundaries projected onto the $(g_0-\vaf)$ planes.

\section{Energetics and The First Law}{\label{sm.sec.energetics}}

\begin{figure*}[ht]
    \centering
        \centerline{\includegraphics[width=17cm]{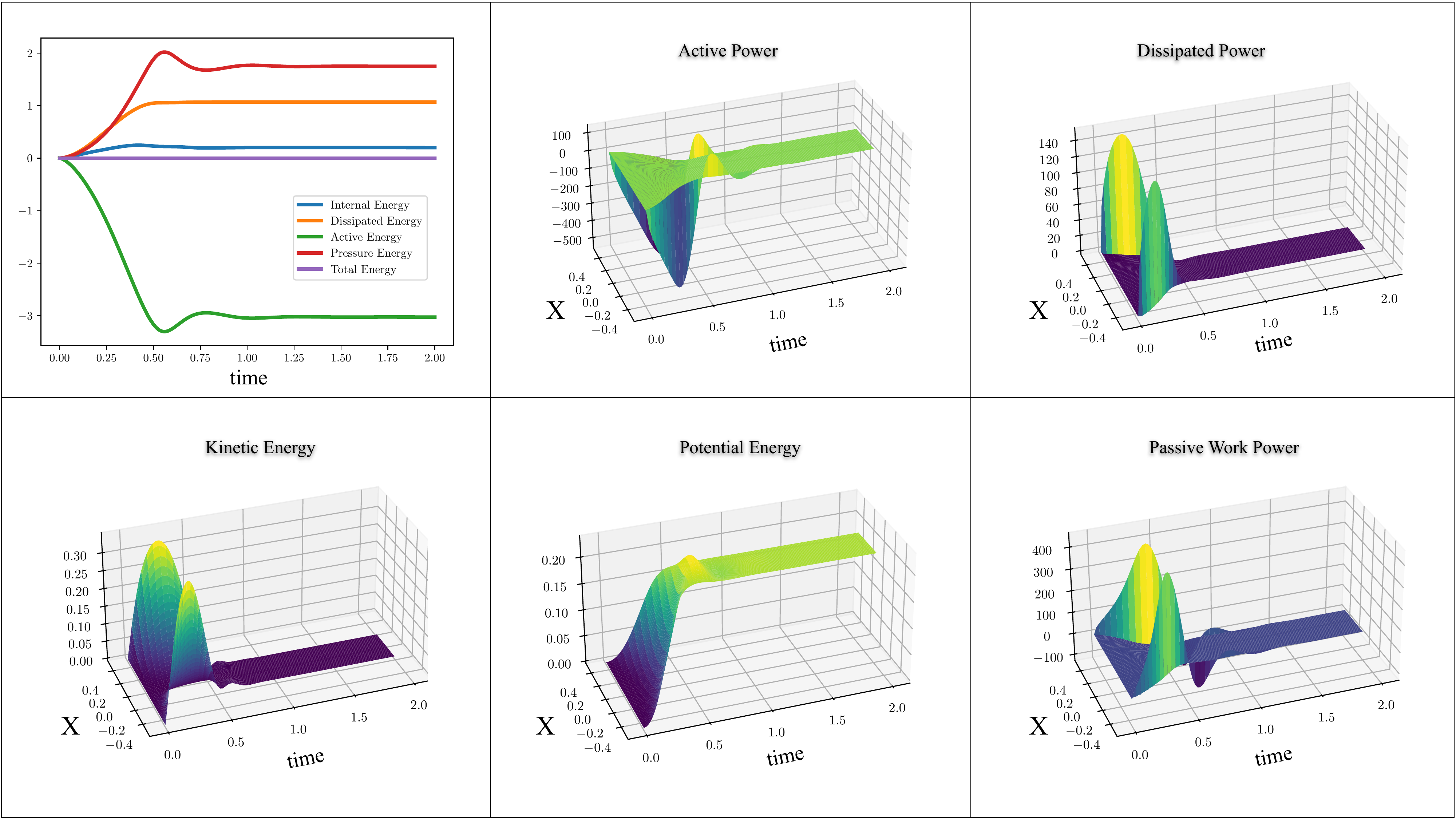}}
            \caption{The figure shows different contributions to the energy functionals, for $\kappa = 1$ and $\gamma = 1$ (drag). The top left panel shows energies and powers (integrated over the spatial domain) as functions of time, indicating that the total energy remains constant (violet curve). Other panels show spatially resolved energies and powers.}
        \label{sm.fig.energetics}
    \vspace{-3mm}
\end{figure*}

In order to investigate the dissipated energy in the viscoelastic system, we note that the first law of thermodynamics can be written in terms of the inexact differentials $\db$ of different energetic contributions:
\begin{align}
    \db E + [\db W_a + \db W_p] + \db Q = 0,
\end{align}
where $E$ is the internal energy of the system, including the kinetic and (elastic) potential contributions. Next $\db W$ is the ``external'' work performed on the system, which includes the active energy injected into the system $\db W_a$, as well as the passive energy stored in the form of positive pressure $\db W_p$.

Finally, $\db Q$ is the dissipated energy in the form of viscous, respectively, and/or drag dissipation, originating from the relative motion inside the material itself and/or relative to the background. While the viscous dissipation is a property of the material, the drag dissipation depends on the environment. Since we are interested in the energy conversion and conservation laws in the active material alone, we deliberately neglect the drag force. However, for the sake of completeness, below we discuss the case of nonzero drag. In summary, and as intuition suggests, the drag force acting along the active region slows down the boundary velocity $v_b$, allowing for transport to be possible for in larger regions of the phase diagram, e.g. for larger velocity of the AF.

In this section we consider an active system of with mass conservation, namely without a source of activated material. In Fig. IV.1, the top left panel shows the total energies as functions of time. The total energy is represented by the violet curve which remains zero throughout the dynamics. Each curve is obtained by integrating over the domain of the system at different time points. In the cases of active energy, dissipated energy and the work done by the internal positive pressure, the cumulative energy change is obtained by integrating the inexact differentials of the corresponding powers.

Other panels show the spatial dependence of different contributions to the total energy. The passive work power is the work performed by the passive component of the positive stress. The density of active power shows extremely peaks at the boundaries and at small times, during which the boundaries start contracting quickly due to the internal contractile stress. Subsequently the density gradient approach zero as the density becomes uniform. Likewise, the dissipated power which originates from viscosity shows peaks at the boundaries because of large velocity gradient. The passive work against contraction also exhibits similar behavior. Therefore, the active energy is to a large extent expended to counteract the positive positive pressure, and to a less extent dissipated via viscosity as well as the drag force.

\section{Simulations and Numerical Methods}{\label{sm.sec.numerics}}
In order to simulate the dynamics of the system, we resort to Lagrangian picture of the flow. Lagrangian as opposed to Eulerian picture is particularly useful when the boundaries of the system move, as in our system. We discretize the initial domain of the system and treat each segment as a Kelvin-Voigt viscoelastic element.

A newly activated element is absorbed into the active region and disconnected from the background. To each segment $n$ we assign the variables $\phi_n,u_n,v_n$, which evolve in time. We denote the vectors using the Dirac's bra-ket notation for clarity. Thus $|u\rangle$ is a vector of length $N$ that represent displacements of the $N$ segments; likewise we have $|\phi\rangle$ for the densities. We denote time step size by $\D{t}$, and the spatial resolution by $\D{X}$; which evolve in time, representing the strain function. The mass of each segment thus equals $\D{m} = \phi_i\D{X}(t=0)$. To simulate the continuum equations we need to define the discrete versions of differential and Laplacian operators. The differential operator is indeed identical to the incidence matrix; a matrix the rows of which represent edges connecting the segments (nodes), and the columns represent segments. For $N$ nodes in 1 dimension, the number of edges is $N-1$, hence a matrix of dimensions $(N-1)\times N$. The entries of the incidence matrix are defined as follows: $\mathcal D_{n,m} = (\delta_{n,m-1}-\delta_{n,m})\D{X}$, where $\delta_{i,j}$ is the Dirac delta function. As such the incidence matrix serves as the differential operator: $\mathcal D|u\rangle = |\bep\rangle\D{X}$. The Laplacian operator is transformed into the discrete Laplace-Beltrami operator $\Delta_{\text{LB}}$, which is obtained using the incidence matrix $\mathcal D$. The Laplace-Beltrami matrix is given by: $\Delta_{\text{LB}} = \mathcal D^{\mathsf{T}}\mathcal D$ ; where $\mathcal D^{\mathsf{T}}$ is the transpose of the differential operator. The discrete equations are as follows:

\begin{subequations}
    \begin{align}
        &|\bep(t)\rangle = \mathcal{D}\,|u(t)\rangle,\\
        &|\phi(t)\rangle = \phi_i/(1+|\bep(t)\rangle),\\
        &|\sigma(t)\rangle = - \alpha_1|\phi(t)\rangle + \alpha_2|\phi^2(t)\rangle,\\
        &|v(t+\D{t})\rangle = |v(t)\rangle + \frac{1}{\D{m}}|f(t)\rangle\D{t},\\
        &|f(t)\rangle = -\eta\,\Delta_{\text{LB}}|v(t)\rangle -\kappa\, \Delta_{\text{LB}}|u(t)\rangle - \mathcal D^{\mathsf{T}}|\sigma(t)\rangle,\\
        &|u(t+\D{t})\rangle = |u(t)\rangle + |v(t)\rangle \D{t} + \frac{1}{2}\frac{1}{\D{m}}|f(t)\rangle\D{t^2}.
    \end{align}
\end{subequations}

We use the above equations to evolve the density, displacement, and velocities. As the AF proceeds, new elements are activated at density $\phi_i$ and gets connected with the adjacent, previously activated, element. 

\subsection{Failure of Transport: Particle's Escape}
In order to investigate the transport or lack thereof, we can use two approaches that can be shown to be equivalent: (1) Explicit simulation of particle's motion attached to the active region; and (2) Simulating the active system alone, finding the density and velocity fields, and using the analytically found conditions for the particle to stay connected, as elaborated in the MT. This condition reads: $v_b/\phi_b\leq g_0/\Gamma$. The reason we are allowed to use either of the above two methods is that the reciprocal interaction of the particle on the field is neglected, namely the field is not affected by the presence of the particle. Therefore, approach (2) is straightforward: we simulate field, and identify regions in the phase diagram which respect the inequality.

For explicit simulation of particle's motion, we need to carefully define the boundary conditions for the active region. Given that the particle's effect is neglected, the field follows a free-boundary condition, which reaches the boundary density $\phi_b$. The force on the particle is always found by calculating the $\nabla\phi_a(X_p)$, i.e. the density gradient at the position of the particle. The gradient is straightforward to calculate everywhere except at the boundary, which is what we care about the most. As argued previously, the boundary provides the local minimum of the potential energy.

The main challenging question in this regard is to determine the conditions under which the particle escapes the potential defined by the active region, which requires defining a ``width'' for the boundary $w$, which is $w \simeq \ell$, where $\ell$ is the particle's linear size. As long as we are concerned with length scales larger than the particle's size, the exact functional form of regularization of the boundary density decay is not important. We can adopt a linear form or a $\tanh$-like function.

The width of the boundary also determines the slope of the potential $\sim g_0\phi_b\ell/w$, which equals the corresponding force exerted on the particle by the field at the boundary. The particle is considered detached if the particle falls behind the boundary's position further than the width of the boundary: $x_b-X_p\geq w$. This is also equivalent to the condition that the particle escapes the potential well of depth $-g_0\ell\phi_b$, if the work performed by the drag force $W_{\text{drag}} = \int^x F_{\text{drag}}\dd x$, over the boundary width, is larger than the potential well: 
\begin{align}
     & W_{\text{drag}} + \uu \geq 0,\\
     & (-\Gamma V_p)(-w) - g_0\ell\phi_b \geq 0,\\
     & \Gamma V_p \geq g_0\phi_b (\ell/w) \simeq g_0\phi_b.
\end{align}
Noting that for the trapped particle $V_p=v_b$, the last equation is identical to that obtained from the force equations. 

\bibliography{refs}

\end{document}